\documentclass[structabstract]{aa}  
\usepackage{float}
\usepackage{hyperref}
\usepackage[T1]{fontenc}
\usepackage{epstopdf}
\usepackage{natbib}
\usepackage{soul}

\usepackage{graphicx}
\usepackage{txfonts}
\usepackage{xcolor}

\begin{document}

   \title{A re-investigation of debris disc halos}

   \author{P. Thebault
          \inst{1}
          \and
          J. Olofsson\inst{2,3,4}
          \and
          Q. Kral\inst{1}
          }
   \institute{LESIA-Observatoire de Paris, UPMC Univ. Paris 06, Univ. Paris-Diderot, France
    \and
    Max Planck Institut f\"ur Astronomie, K\"onigstuhl 17, 69117 Heidelberg, Germany
    \and
    Instituto de F\'isica y Astronom\'ia, Facultad de Ciencias, Universidad de Valpara\'iso, Av. Gran Breta\~na 1111, Playa Ancha, Valpara\'iso, Chile
    \and
    N\'ucleo Milenio Formaci\'on Planetaria - NPF, Universidad de Valpara\'iso, Av. Gran Breta\~na 1111, Valpara\'iso, Chile
             }  
\offprints{P. Thebault} \mail{philippe.thebault@obspm.fr}
\date{Received ; accepted } \titlerunning{halos of debris discs}
\authorrunning{Thebault, Olofsson \& Kral}

%

%

% \abstract{}{}{}{}{} 
% 5 {} token are mandatory
 
  \abstract
   {Scattered-light images reveal that a significant fraction of debris discs consist of a bright ring beyond which extends a wide halo. Such a halo is expected and should be made of small grains collisionally produced in the ring of parent bodies (PB) and pushed on high-eccentricity orbits by radiation pressure. It has been shown that, under several simplifying assumptions, the surface brightness (SB) of this halo should radially decrease as $r^{-3.5}$ in scattered light}
   {We aim to revisit the halo phenomenon and focus on two so far unexplored issues: 1) How the  unavoidable presence of small unbound grains, non-isotropic scattering phase functions (SPF) and finite instrument resolution affect scattered light SB profiles, and 2) How the halo phenomenon manifests itself at longer wavelengths in thermal emission, both on resolved images and on system-integrated SEDs}    
   {We use a collisional evolution code to estimate the size-dependent spatial distribution of grains in a belt+halo system at steady state. We use the GRaTeR radiative transfer code to derive synthetic images in scattered-light and thermal emission, as well as SEDs.}
   {We find that unbound grains account for a significant fraction of the halo's luminosity in scattered light, and can significantly flatten the SB radial profile for the densest and brightest discs. Because halos are strongly size-segregated with radial distance, realistic size-dependent SPFs also have an affect, resulting here again in shallower SB profiles. For edge-on discs, non-resolving the vertical profile can also significantly flatten the projected SB profile. We show that roughly half of the observationally-derived halo profiles found in the literature are compatible with our new results, and that roughly half of the remaining systems are probably shaped by additional processes (planets, stellar companions, etc.). We also propose that, in future observational studies, the characteristics of PB belt and halos should be fitted separately. In thermal emission, we find that wide halos should remain detectable up to the far-IR and that, with the exception of the $\sim 8-15\mu$m domain, halos account for more than half of the system's total flux up to $\lambda\sim80-90\mu$m. The halo's contribution strongly decreases in the sub-mm to mm but still represents a few percents of the system's luminosity at $\lambda\sim 1$mm. For unresolved systems, the presence of a halo can also affect the determination of the disc's radius from its SED. }
   {}

   \maketitle
%
%________________________________________________________________

\section{Introduction} \label{intro}

Circumstellar debris discs have been detected around a significant fraction (15 to $30\%$) of main sequence stars \citep{hughes2018}. These discs are thought to result from the continuous collisional grinding of leftovers from the planet-formation process, of which only the tail end (dust particles) is detectable as a photometric infra-red (IR) excess \citep{wyatt2008}. 
An increasing number of these discs have also been imaged, most commonly by scattered light observations in the visible and near-infrared, but also in thermal emission in the mid-to-far IR to millimetre wavelength domain \footnote{See the regularly updated database of resolved discs available at 
\url{https://www.astro.uni-jena.de/index.php/theory/catalog-of-resolved-debris-disks.html}}. These images have revealed a variety of structures, such as clumps, warps, spirals or arcs, which have been interpreted as being sculpted by gravitational perturbations of (usually unseen) planets \citep[e.g.][]{augereau2001,wyatt2006,theb2012} or companion stars \citep{theb2021}, or to be the result of violent and transient collisional events \citep{jackson2014,kral2015,theb2018}.

Apart from the aforementioned spatial structures, most imaged discs share a common feature, which is that the dust surface density is not uniformly decreasing (or increasing) with stellar distance, but peaks at a given radial location, thus creating ring-like features. This belt-like configuration is so ubiquitous that it has been suggested that "debris rings" should be a more appropriate denomination for the debris discs phenomenon \citep{strubbe2006}. In many cases, scattered-light images also show a dimmer but radially extended halo of dust beyond the location of this main ring or belt \citep{hughes2018}. The analytical and numerical studies of \citet[][ hereafter STCH]{strubbe2006} and \citet[][ hereafter TBWU]{theb2008} have shown that such halos are in fact expected beyond collisionally evolving belts of parent bodies (PB). These halos should be made of small, typically 1-10$\mu$m sized grains that are collisionally produced within the main ring and then placed on high-eccentricity orbits by radiation pressure or stellar wind. For systems dense enough for the Poynting-Robertson (PR) effect to be negligible with respect to collisional evolution, the vertical optical depth $\tau$ in the outer halo naturally tends towards a radial profile in $\tau\propto r^{-1.5}$. At wavelengths dominated by scattered light, this translates into a surface brightness (SB) profile that decreases as $\propto r^{-3.5}$. This SB $\propto r^{-3.5}$ slope also holds for the projected mid-plane surface brightness profile of discs seen edge-on, provided that their vertical scale height $z$ is $\propto r$ \citep{strubbe2006}. Note, however, that these results are valid under several simplifying assumptions, notably isotropic scattering and the absence of small unbound grains that are blown out by radiation pressure or stellar wind.

Thanks to cutting-edge instruments such as SPHERE or GPI \citep{beuzit2019, macintosh2014}, recent years have seen an exponential increase of the number of systems for which the radial profile of halos has been constrained \citep{adam2021}. The canonical radial profiles of STCH and TBWU can then be used as benchmarks to judge as to whether or not the outer edge of a debris ring is "natural" or, on the contrary, shaped by additional mechanisms (outer planet, interaction with gas, companion star, etc.). However, the quantity whose radial profile is constrained by observations can differ from one study to the other. In some cases, it is the surface brightness (SB) that is retrieved, which can either be the (deprojected) stellocentric SB measured along a radial cut \citep{schneider2018} or the projected SB in the case of discs seen edge-on \citep{golimowski2006}. In other cases it is the underlying grain number density profile that is estimated, usually by fitting the observed resolved images with radiative transfer models, such as the GRaTer code \citep{augereau1999}, in which several free parameters (parent belt location and width, radial slope in the halo, scattering phase function, etc.) are explored and adjusted \citep{choquet2018,bhowmik2019,perrot2019}. The radial density profile that is estimated can either be the surface density $\sigma$ or the volumic number density $n$.

\begin{table*}[!t]
\begin{center}
\begin{tabular}{l l r c c c c c c l }
\hline\hline

Name & $f_d$=($L_{disc}/L_*$)$_{\rm{IR}}$ & $r_0$(au) & $r_{\rm{max}}(au)$ & SB & $n$ & $\sigma$ & $i$(deg) & Reference\\
 & &  &  &(-3.5)&(-2.5)&(-1.5)&  & \\
\hline
HD139664 & $9\times10^{-5}$ & 83.1 & 109.4 && -2.5 & & 87 & \citet{kalas2006} \\
HD141943$^{d}$ & $1.2\times 10^{-4}$& 115 & 145 && -4 && 87 & \cite{boccaletti2019}\\
HD160305 & $1.2\times10^{-4}$ & 86 & 106 && -7.1 && 82.3 & \citet{perrot2019}\\
HD202628 & $1.4\times10^{-4}$ & 175 & 200;250 & -4.5;-3.9/-12.9;-2.8 &&& 61 & \citet{schneider2016}\\
HIP67497 & $1.4\times10^{-4}$ & 60.7 & 70 && -8.5 or -1.3$^{c}$ && 80 & \citet{bonnefoy2017}\\
HD53143 & $2.5\times10^{-4}$ & 82 & 110 & -3 & -1 && 45 & \citet{kalas2006}\\
AUMic & $3.9\times10^{-4}$ & 33 & 60 & -3.8 & & & 90 & \citet{fitzgerald2007}\\
HD15115 & $5\times10^{-4}$ & 98 & 340/580 & -4.7/-3 & -5.5/-3.5 & & 86 & \citet{engler2019}\\
HD143675 & $5.6\times10^{-4}$ & 48.1 & 55.6 && -3 && 87.2 & \citet{esposito2020} \\
HD192758 & $5.7\times10^{-4}$ & 95 & 170(?) && -2 && 58 & \citet{choquet2018}\\
HD104860 & $6.3\times10^{-4}$ & 114 & 160 && -3.9 && 54 & \citet{choquet2018} \\
HD172555 & $7.2\times10^{-4}$ & 10.9 & 15 && -9.8 && 103.5 & \citet{engler2018}\\
HD157687 & $7.9\times10^{-4}$ & 78.9 & 211 && -2.2 && 68.3 & \citet{millar2016}\\
Fomalhaut & $9\times10^{-4}$ & 154 & 209 & -3.3 & & & 66 & \citet{kalas2013} \\ 
49Cet & $9\times10^{-4}$ & 129 & 270 & & -2.1 & & 79 & \citet{choquet2017}\\
HD107146 & $1.2\times10^{-3}$ & 135.5 & 185 &-4.8&& -2.8 & 18 & \citet{ardila2004} \\
HD145560 & $1.27\times10^{-3}$ & 85.3 & 108 && -3 && 44 & \citet{esposito2020} \\
HD106906$^{b}$ & $1.3\times10^{-3}$ & 67.6 & 110 && -4 && 85 & \citet{lagrange2016} \\
HD35841 & $1.3\times10^{-3}$ & 56 & 110 & -3.55 &&& 84.9 & \citet{esposito2018}\\
HD191089$^{d}$ &  $1.4\times10^{-3}$ & 45 & 60;400 & -6.1;-2.68 &&& 59 & \citet{ren2019}\\
HD131835 & $1.4\times10^{-3}$ & 96 & 140 && -2.3 && 75.1 & \citet{feldt2017}\\
TWA 7$^{a}$  & $1.7\times10^{-3}$& 25 & 70 & & -1.5 & & 13 & \citet{olofsson2018}\\
HD115600 & $1.7\times10^{-3}$ & 48 & 60 && -7.5 && 79.5 & \citet{currie2015}\\
HD181327 &  $2\times10^{-3}$ & 90 & 150;250 &&& -3.7;-1.7 & 32 & \citet{stark2014}\\
HD15745 & $2.2\times10^{-3}$ & 165 & 450 & -3.5 &&& 65 & \citet{kalas2007}\\
$\beta$ Pic & $2.4\times10^{-3}$ & 127 & 193;258 & -4;-3/-4.5;-3.5 & & & 90 & \citet{golimowski2006}\\
HD117214 & $2.67\times10^{-3}$ & 60.2 & 97.2 && -4.5 && 71 & \citet{esposito2020} \\
HD32297$^{e}$ & $2.7\times10^{-3}$ & 130 & 400 &-3.2 or -5.3 & -4 or -6 && 90 & \citet{bhowmik2019}\\
HD114082 & $3\times10^{-3}$ & 29.6 & ? &&& -3.9 & 83 & \citet{wahhaj2016} \\
HD61005 & $3\times10^{-3}$ & 60 & ? && -2.7 && 84.1 & \citet{olofsson2016}\\

HD36546 & $4\times10^{-3}$ & 82 & 110 &-2.5&-1.5 && 78.9 & \citet{lawson2021}\\
HD111161 & $4.2\times10^{-3}$ & 72.4 & 98.1 && -3 && 62 & \citet{esposito2020} \\
HD156623 & $4.3\times10^{-3}$ & 80 & 94 && -3.5 && 30 & \citet{esposito2020} \\
HD121617 & $4.8\times10^{-3}$ & 78.3 & 128 && -5.6 && 43.1 & \citet{perrot2023}\\
HR4796 & $5\times10^{-3}$ & 81.4 & 330/123;330 & -5.1/-7.8;-3.8 & & & 76 & \citet{schneider2018}\\
HD129590 & $5\times10^{-3}$ & 59.3 & 141 && -1.3 && 74.6 & \citet{matthews2017}\\
HIP79977 & $5.21\times10^{-3}$ & 73 & 225 && -2.5 && 84.6 & \citet{engler2017}\\
\hline
\end{tabular}
\end{center}
\caption{ Debris discs, ranked by increasing fractional luminosity $f_d$, with resolved halos in scattered light, for which a power-law fit of the outer radial profile of either the surface brightness (SB), the volumic number density $n$ or the surface density $\sigma$ is available in the literature. $r_0$ is the location of the density peak in the main belt, $r_{max}$ the radial distance out to which the radial slope of the halo has been constrained. When two different profiles have been obtained for opposite sides of the disc, a "/" is inserted between the two fits. When the fitted radial slope has the form of a broken power-law, two $r_{max}$ values are given, separated by a ";", the first value being the distance at which the slope changes. Likewise, two radial index values are also given, also separated by a ";". The values in  parenthesis given at the top of the SB, $n$ and $\sigma$ columns correspond to the expected radial index values in the simplified scenario of STCH and TBWU. \\ Notes: (a) There is a tentative detection of a second ring at 52au that could be the cause the apparently shallow profile (b) Disc around a central binary. Imaged planet at a projected separation of 650au (c) The 2 indices correspond to 2 separate fits: one with a single belt and one with a double belt \citep{bonnefoy2017} (d) (NZ Lup) The slope given here is that beyond the outer ring in the "2 belt" scenario considered \citep{boccaletti2019} (e) The values for the SB slopes are derived by eye from Fig.10 of \cite{bhowmik2019} }
\label{tab:slopes}
\end{table*}

\subsection{A coherent census of observationally-derived halo profiles} \label{census}

The fact that, depending on the profile-fitting approach, the notion of radial profile can refer to three distinct parameters (SB, $\sigma$ and $n$) sometimes lead to some confusion when comparing different discs or when comparing observations to canonical theoretical profiles. An additional problem is that the theoretical results of STCH and TBWU constrain the system's optical depth $\tau$ (or geometrical cross section), and not the particle number densities $\sigma$ and $n$. If the disc was only made of identical particles, then the radial profiles of $\tau$ and $\sigma$ (and also $n\times r$ for non-flared discs) should be equivalent \footnote{ because the variations of the geometrical cross-section should directly follow the variations of the number of particles}, but it is not true in the present case because there is a very strong size segregation as a function of radial distance in the halo \citep{theb2014}. Schematically, at a given stellar distance $r$ the cross section is indeed dominated by grains produced in the main ring that have their apoastron $Q=r$. Since the value of $Q$ is imposed by the size-dependent mechanism that is radiation pressure (or, for M stars, stellar wind), it follows that, at a distance $r$, $\tau$ is dominated by grains of size $s$ such that
\begin{equation}
    \beta(s) = \frac{1}{2} \left( 1-\frac{r_0}{r}\right)
\label{equ:domsize}
\end{equation}
where $r_0$ is the location of the main belt and $\beta(s)$ is the ratio between the radiation pressure and stellar gravity (or stellar wind) forces for the size $s$.

In contrast, the number density profiles derived in most GRaTeR-type fits assume that the particle size distribution is the same everywhere in the system, which clearly cannot apply to the size-segregated halo. This means that neither $\sigma$ nor $n$ derived this way correspond to actual number densities (neither that of the global population nor that of a given size range). Nevertheless, for fits performed on images at short scattered-light wavelengths, the radial \emph{dependence} of the GRaTeR-derived $\sigma$ parameter should, to a first approximation, match that of $\tau$ as long as grains in the halo contribute to the flux proportionally to their cross section. As will be shown in Sec.\ref{results}, this assumption does not hold if small unbound grains have a significant contribution and if the scattering phase function is size-dependent, but before exploring these important issues, let us present in Tab.\ref{tab:slopes} the first coherent census of all systems with observationally-constrained halo profiles for which we specify which quantity it is (SB, $n$ or $\sigma$) that has had its radial profile constrained. 
It can be noted that the radial index $\Gamma$ of the surface density profile has been directly estimated for only 3 systems, and that the most commonly constrained slope is $\alpha_{out}$ for the volumic density $n$ and, to a lesser extent, $\gamma_{out}$ for the stellocentric SB profile.

In most studies, the comparison to the theoretical results of STCH and THWU
is done by assuming that the profiles of $\tau$ and $\sigma$ are identical (see previous paragraph) and that the indexes $\Gamma$, $\alpha_{out}$ and $\gamma_{out}$ are related through the relations
\begin{equation}
    \Gamma = \gamma_{out} + 2
\label{gamma}
\end{equation}
and
\begin{equation}
    \Gamma = \alpha_{out} + \delta
\label{alpha}
\end{equation}
where $\delta$ is the index of the radial profile of the halo's scale height, which is usually assumed to be equal to 1 \citep{thalmann2013,ren2019}. For the specific case of edge-on seen discs, we have $\Gamma = \gamma_{out}+1+\delta$, which reduces to the same $\Gamma = \gamma_{out} + 2$ relation for $\delta=1$ \citep{strubbe2006}.

Note, however, that these relations are in principle only valid in an idealized case with isotropic scattering as well as for grains that are large enough for their contribution to the flux to be proportional to their geometrical cross section, and, in the case of edge-on discs, for an infinite instrumental resolution in the vertical direction.

\subsection{halos in thermal emission}

All halos listed in Tab.\ref{tab:slopes} have been imaged in scattered light and, more generally, the halo phenomenon has so far mostly been investigated in the visible or near-IR and not in thermal emission. There are two main reasons for that. The first one is theoretical and is that halos are supposed to be made of small grains not much bigger than the radiation-pressure blowout size $s_{blow}$, which are not expected to contribute at long wavelengths where they are poor emitters. The second reason is observational: high-end instruments in the visible or near-IR are by far those offering the best spatial resolution allowing to resolve the structures of halos. 

However, the recent discovery of two extended halos around HD32227 and HD61005 with ALMA in the millimetre \citep{macgregor2018} has shed new light on the matter. Based on order-of-magnitudes flux-to-mass conversions, \cite{macgregor2018} estimated that these halos were mostly made of large mm-sized grains, which are not supposed to populate halos according to the canonical scenario of STCH and TBWU. To explain this puzzling presence, these authors explored some additional mechanisms, such as planetary or stellar perturbations, interaction with the ISM, or the aftermath of a large planetesimal breakup, but could not find a fully satisfying scenario. 
More recently \cite{olofsson2022b} revisited the modeling of the ALMA observations of HD\,32297 (jointly with SPHERE polarimetric data) and found that one cannot rule out the possibility that these mm-halos are mostly made of smaller micron-sized grains after all. They showed that an overdensity of micron-grains, as expected because of their longer collisional lifetimes, could in principle compensate for their lower emissivity. However, using the simplified STCH and TBWU relations, they found that the flux due to these grains could only account to up to $\sim30\,\%$ of the measured ALMA levels, but more sophisticated grain-distribution modelling might change these preliminary results and this crucial issue thus remain an open question.
Additional halo-detections in the mm should be expected in the near future, as ALMA is slowly catching up with the resolution of near-IR instruments such as SPHERE and GPI, and high angular resolution projects are currently being executed (e.g. ARKS large program, PI Marino). In addition, new ground- and spaced-based facilities (e.g., VLT/ERIS and JWST) can now observe debris disks in the mid-IR with unprecedented precision, which might potentially allow to image halos at these wavelengths, underlining the need for a more comprehensive study of their detectability in thermal emission.

\subsection{paper outline}

In the light of these pending issues and recent new developments, we undertake a thorough numerical re-investigation of the halo phenomenon. We will focus on two main problems. Firstly, how the scattered light radial profiles derived by STCH and TBWU might be affected by taking into account the effect of small unbound grains, size-dependent scattering phase functions and, for edge-on discs, instrument resolution in the vertical direction. Secondly, we explore how the belt+halo phenomenon manifests itself at longer wavelengths, both in terms of the system's radial brightness profile as well as of the halo's imprint on the integrated Spectral Energy Distribution (SED).

We briefly present in Sec.\ref{model} the collisional evolution code that will serve as a basis in our study as well as the typical belt+halo setup that we will explore. Secs.3.1 and 3.2 present the results of our numerical exploration regarding radial profiles in scattered light. Results regarding profiles at longer wavelength as well as system-integrated SEDs are presented in Sec.3.3. We discuss the implications of our results in Sec.4 and conclude in Sec.5.

\section{Model}\label{model}

\subsection{Basic principle}

Our numerical investigation will use the tried and tested particle-in-a-box collisional model initially created by \cite{theb2003}, and constantly improved during the past 2 decades. As with similar codes, such as the ACE code of the Jena group \citep{krivov2005}, particles are sorted into logarithmic size bins, whose populations are evolved following estimates of their mutual impact rates and collisional outcomes. This code has a 1D spatial resolution and is divided into radially concentric annuli.
We use the latest version of the code, as described by \cite{theb2019}, for which collisional rates are estimated by separate deterministic $N$-body simulations taking into account the effect of stellar gravity and radiation pressure \footnote{under the assumption that no additional perturbing body is present}, and with a more realistic collisional prescription for small particles in the $\lesssim 1$mm range taken from the laboratory experiments by \cite{nagaoka2014}. We refer the reader to \cite{theb2007} and \cite{theb2019} for a detailed description of the code.

To derive surface brightness profiles at different wavelengths as well as SEDs we use the (also tried and tested) GRaTeR radiative-transfer package \citep{augereau1999}.

\subsection{Setup}\label{setup}

\begin{table}[h]
\caption{Numerical setup}
\centering
\begin{tabular}{lc}
\hline\hline

Stellar-type & A6V \\
Stellar luminosity & 8.7$L_{\odot}$ \\
Stellar Temperature & 8052K \\
\hline
Material & Compact astrosilicate \\
Bulk density & 2.7g.cm$^{-3}$ \\
Blow-out size ($s_\mathrm{\rm{blow}}$) & $2.5\,\mu$m \\
Minimum particle size & $ 0.025\,\mu$m  \\
Maximum particle size   & 20\,km  \\
Dynamical excitation in the PB belt & <$e_0$>=2<$i_0$>=0.05\\
Radial extent of the PB belt & 50-66\,au \\
fractional luminosity at steady state & $f_{d}\sim 8\times10^{-4}$ (nominal)\\
 & $f_{d}\sim 4\times10^{-3}$ (bright disc)\\
\hline
\end{tabular}
\label{setupt}
\end{table}

For the sake of readability of our results and in order to avoid numerically-expensive parameter explorations, we consider a reference "nominal" setup, chosen as to be the most representative of observed belt+halo systems. The considered setup is that of a narrow belt of parent bodies, extending from 50 to 66\,au, thus centered at $r_0=58\,$au and of width $\Delta r_{0}=16\,$au, where all the mass is initially confined. 
We consider 119 log size bins, between $s_{max}=20$km and $s_{min}=0.025\mu$m. Large parent bodies in the belt (large enough for radiation pressure effect to be negligible) are assumed to be located on orbits of average eccentricity <$e_0$>=0.05 and inclination <$i_0$>=<$e_0$>/2=0.025, which are typical values for debris-producing discs \citep[e.g.,][]{theb2009}. Regarding grain composition, we consider the generic case of compact astrosilicates \citep{draine2003}.
As for the central star, we consider an A6V stellar type, identical to the archetypal $\beta$ Pictoris case. For this stellar type, the blow-out size $s_{blow}$ due to radiation pressure is of the order of $2\mu$m, which is much larger than our $s_{min}$, meaning that our code takes into account the potentially important effect of unbound grains. 

Another important parameter is the level of collisional activity within the disc, which can be, to a first order, parameterized by $f_d$, the disc's fractional luminosity in IR. We consider as a reference case a system with $f_d=8\times10^{-4}$, which is approximately the average value for the resolved-halos systems presented in Tab.\ref{tab:slopes}. In the spirit of \cite{theb2019}, we also consider a "very collisionally active" case with $f_d=4\times10^{-3}$, corresponding to the brightest discs in our census (such as HD32297 or HD129590).
As explained in \cite{theb2019}, in practice we always start with discs whose initial masses are expected to correspond to $f_d$ larger than the ones we are aiming for, and then let the systems collisionally evolve until (1) the shape of the particle size distribution (PSD) no longer changes (steady state), and (2) $f_d$ has decreased to the desired value.

All main parameters for this nominal setup are summarized in Tab.\ref{setupt}.

\section{Results}\label{results}

\subsection{Radial profile in scattered light}

\begin{figure}
\includegraphics[scale=0.5]{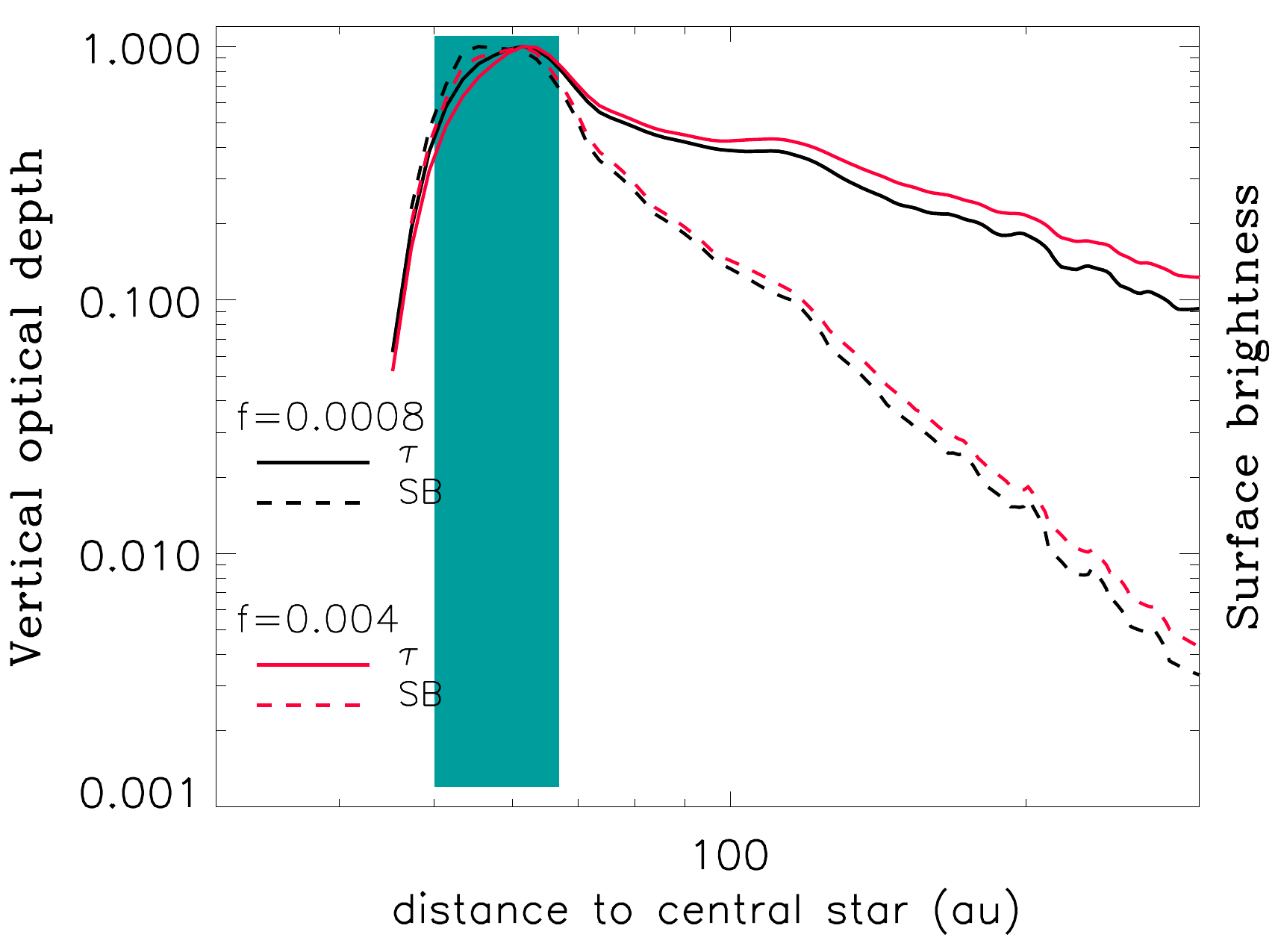}
\caption[]{Normalized radial profiles of the vertical optical depth ($\tau$) and the surface brightness (SB), in scattered light ($\lambda=0.8\mu$m), for the nominal setup presented in Tab.\ref{setupt}, as well as for a "very bright" disc case (fractional luminosity $f_d=4\times10^{-3}$). The blue area marks the radial extent of the parent body belt. Scattering is here assumed to be isotropic}
\label{optflux}
\end{figure}

Fig.\ref{optflux} presents the radial profile, at steady state, of both $\tau$ and the SB in scattered light, for the nominal case as well as for the "bright disc" case, assuming isotropic scattering when deriving the SB.
Strictly speaking, the displayed SB(r) profiles correspond to a disc seen face-on ($i=0$), but they should also be valid for any other viewing configuration as long as the disc is not seen edge-on, i.e., as long that, at a given position along any radial cut of the observed disc, there are only grains from a given stellocentric distance that contribute to the flux. To a first approximation, this is true if the disc's opening angle $\psi$ is less than $90-i$. For these non-face-on cases, the displayed SB profiles are the ones that should be obtained for the deprojected stellocentric luminosity profile along any radial cut. 

\begin{figure}
\includegraphics[scale=0.5]{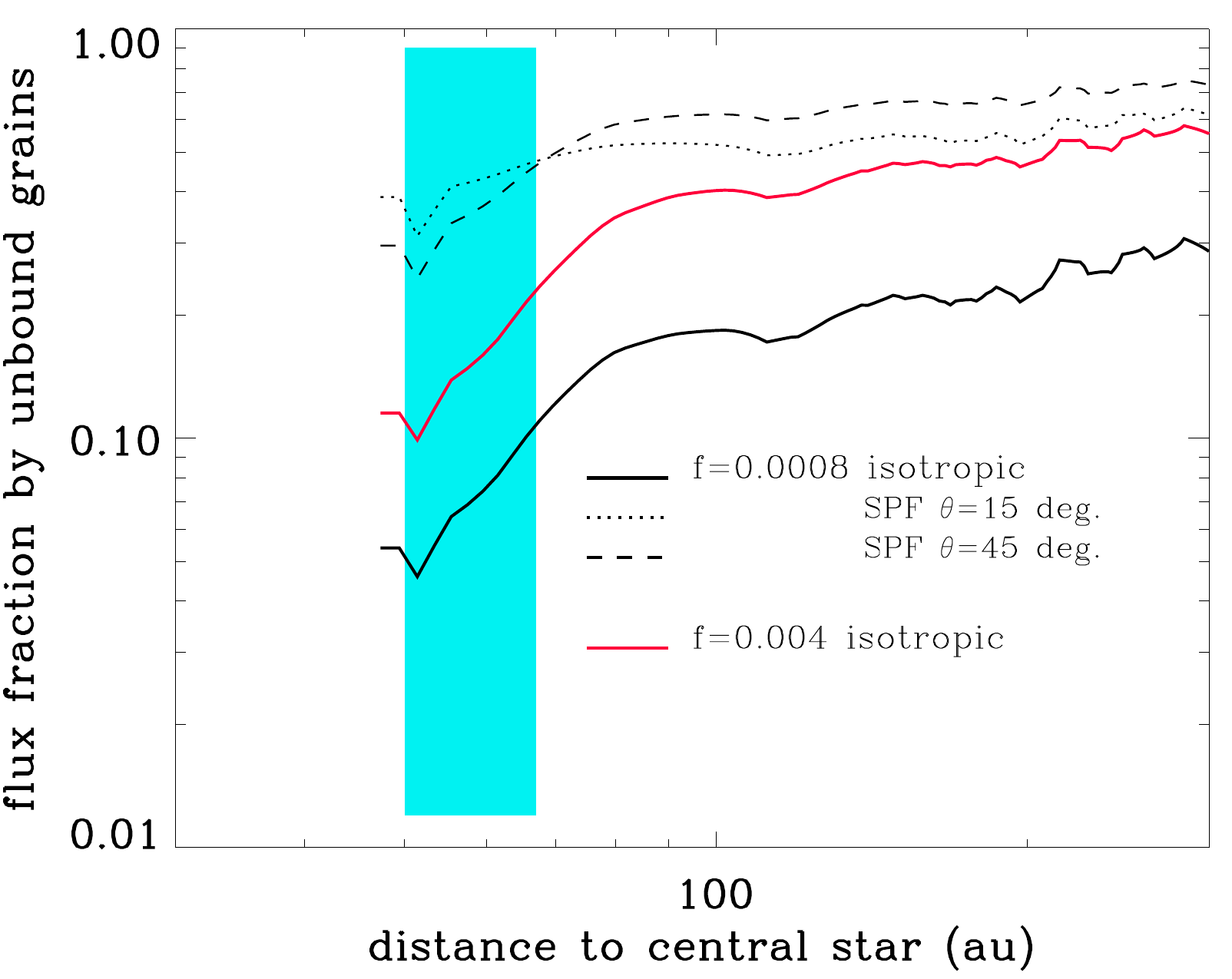}
\caption[]{Radial dependence of the fraction of the flux density, at 0.8$\mu$m in scattered light, that is due to unbound grains ($s<s_{blow}$). Results are shown for the nominal and "bright disc" cases, as well as for anisotropic scattering, at 2 different angles, using the distribution of hollow spheres model (DHS, \citealp{Min2005}) for the scattering phase function (SPF)}
\label{unbcurve}
\end{figure}

The average radial slope that is asymptotically reached by the optical depth profiles is -1.48 for the nominal case ($f_d=8\times10^{-4}$), which is very close to the theoretical value of -1.5, and the average slope of the \emph{SB(r)} profile is -3.42, again very close to the canonical value of -3.5 when applying the relation $\Gamma = \gamma_{out} + 2$. The  optical depth profiles are, however, shallower for the highly collisional bright disc case ($f_d=4\times10^{-3}$), with a radial index of -1.21 for $\tau(r)$ instead of -1.48. This is mainly due to the fact that, for such high-$f_d$ cases, unbound grains ($s<s_{blow}$) begin to significantly contribute to the geometrical cross section, the proportion of unbound grains being directly proportionnal to the level of collisional activity in the system \citep[see][]{theb2019}. 
The effect of unbound grains is less pronounced on the flux, because their smaller size, especially the ones in the $\lesssim 0.1\mu$m range, makes them less efficient scatterers. The slope of \emph{SB(r)} in the "very bright" case is still, however, slightly smaller than for the $f_d=8\times10^{-4}$ case, with a radial index of -3.25 instead of -3.42.
In this $f_d=4\times10^{-3}$ run, unbound grains even dominate the flux in the halo beyond $\sim200\,$au (Fig.\ref{unbcurve}). The reason why these $s<s_{blow}$ dust particles will tend to flatten the radial profile is that their distribution should follow $\tau\propto r^{-1}$ instead of $\propto r^{-1.5}$ for bound grains (see section 3.3 in STCH). This flattening effect is less visible for the reference $f_d=8\times10^{-4}$ case, but the contribution of unbound grains to the flux still exceeds $10\%$ everywhere in the halo (see black solid line in Fig.\ref{unbcurve}).

Note that, as already noted by \cite{theb2012}, the asymptotic $\tau$ and \emph{SB} slopes are not reached immediately beyond the PB belt, but after a transition region where the density and flux drop more abruptly. This is because, right outside the PB belt, there is a sudden transition from a ring where all particle sizes are present to an outer region only populated by small, radiation-pressure affected grains. If we define the limiting size $s_{PR}$ for radiation-pressure-affected grains by the criterion 
\begin{equation}
    e(s_{PR})=\frac{\beta(s_{PR})}{1-\beta(s_{PR})}=e_0
\end{equation}
then, for $e_0=0.05$, we get $s_{PR}\sim11 s_{blow}$. For a standard $dN\propto s^{-3.5}$ size distribution, the absence of all $s>s_{PR}$ grains results in a drop of geometrical cross section of $\sim30\%$, which is roughly what we observe on Fig.\ref{optflux}. The width of this transition region at the outer edge of the PB belt is $\sim0.15r_0$ for both the nominal and high-$f_d$ cases.

\begin{figure}
\includegraphics[scale=0.5]{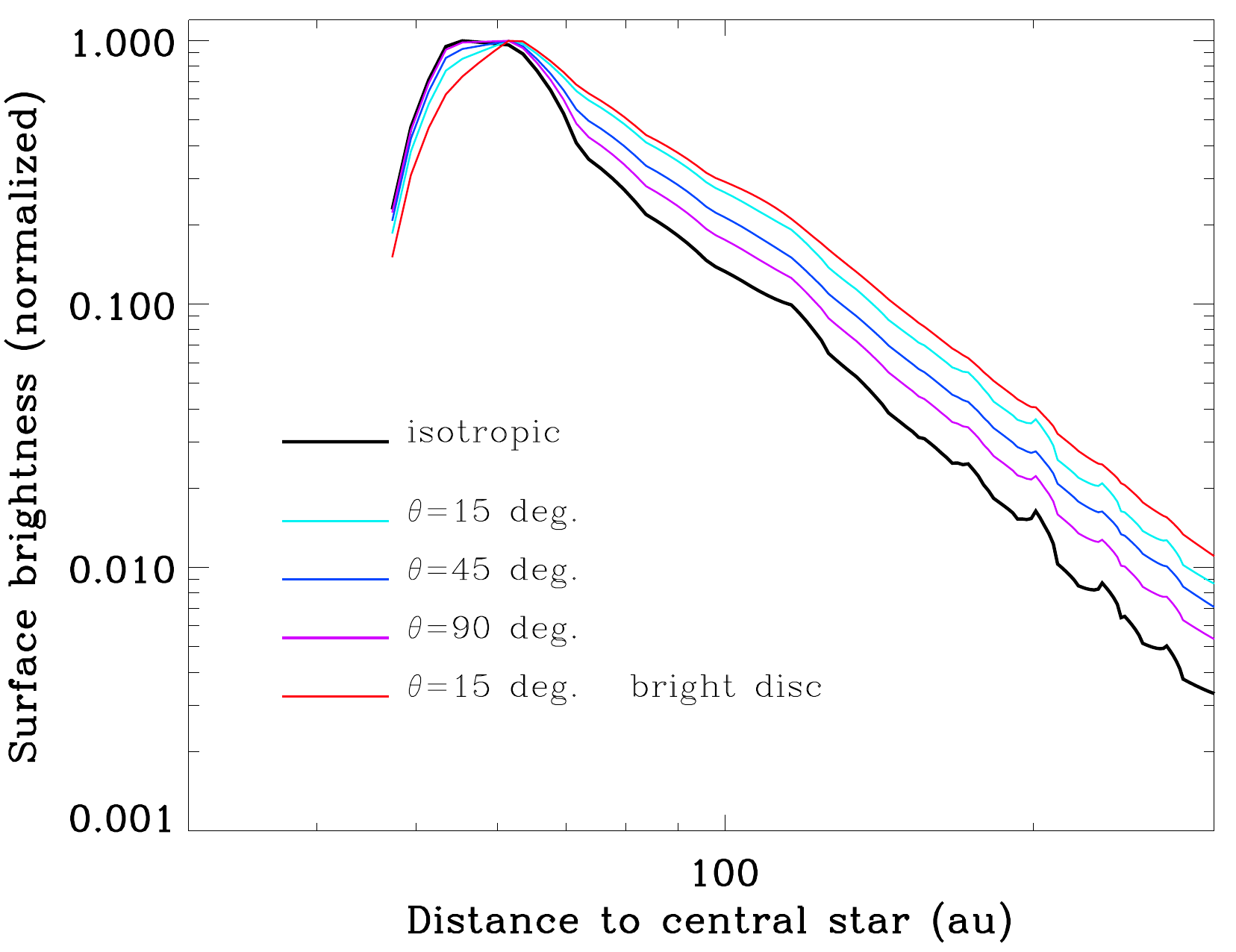}
\caption[]{Radial profile of the normalized surface brightness obtained using the DHS scattering phase function prescription for 3 different scattering angles }
\label{profspf}
\end{figure}

\begin{figure}
\includegraphics[scale=0.5]{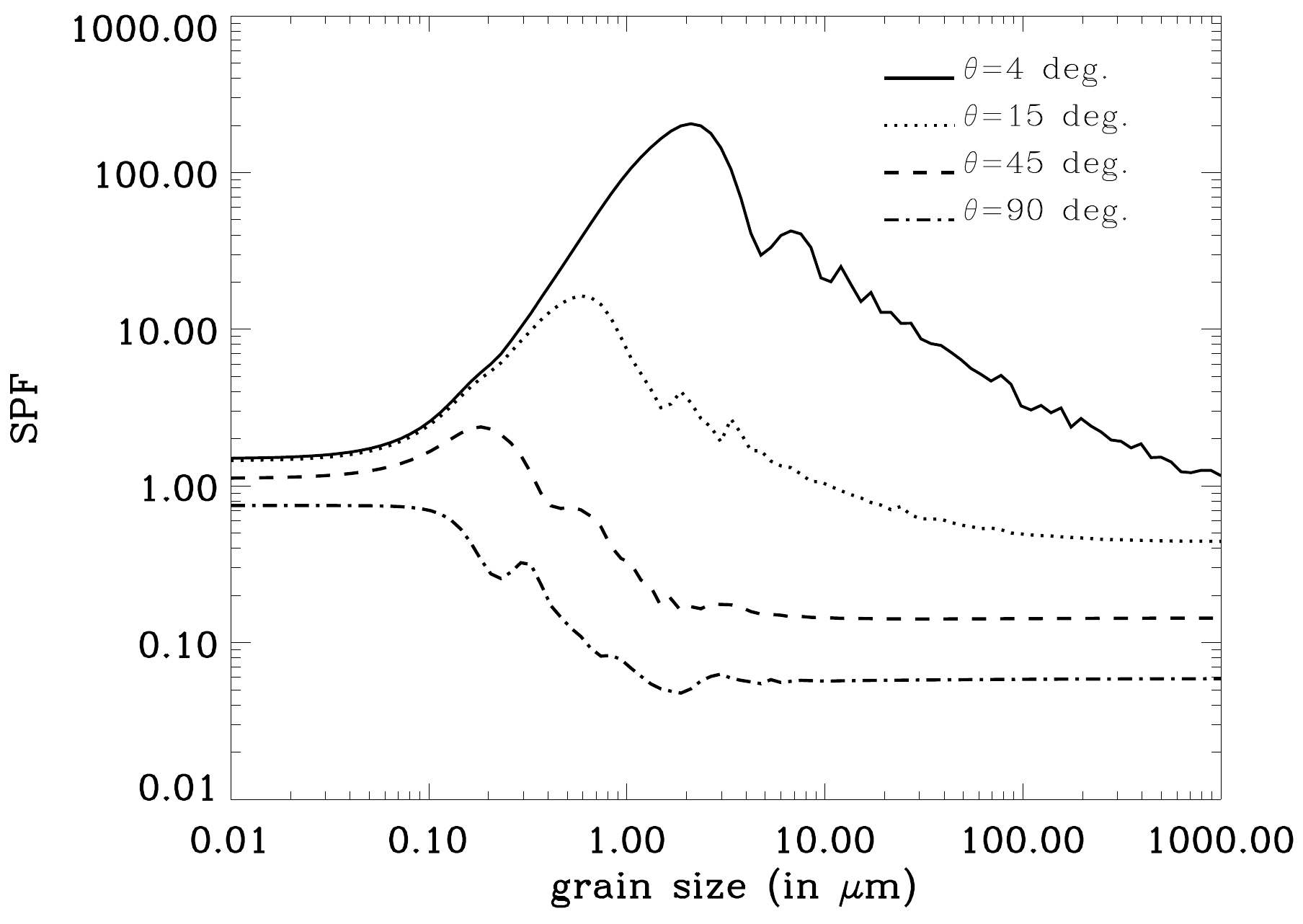}
\caption[]{Size dependence of the DHS scattering phase function, for 4 different scattering angles}
\label{DHS}
\end{figure}

The radial SB slopes of Fig.\ref{optflux} have been derived by implicitly assuming isotropic scattering. Because the viewing angle should not vary along a given radial cut, they are in principle independent of the scattering phase function (SPF) provided that the SPF does not depend on stellocentric distance. This would, for instance, be the case for a standard Henyey-Greenstein SPF prescription \citep{henyey1941} with a constant, size-averaged $g$ parameter. Nevertheless, this assumption might not hold for halos, where the strong grain-size segregation as a function of $r$, coupled to the fact that the SPF is expected to vary with grain size, should result in a radial variation of the phase function. We thus explore this effect further by including a more realistic SPF in the models. The simplest option would be to use the Mie theory (\citealp{Mie1908}), for which the dust particles are assumed to be compact spheres. However, several studies (e.g., \citealp{Rodigas2015}, \citealp{Milli2019}, \citealp{Chen2020}, \citealp{Arriaga2020}) have demonstrated that the assumption of spherical grains does not really hold when trying to reproduce observations of debris disks. Instead, we here use the SPF computed using the distribution of hollow spheres model (DHS, \citealp{Min2005}). For each grain size, the optical properties (e.g., scattering efficiencies, SPF) are obtained by averaging over a distribution of shapes. As discussed in \citet{Min2016}, this model is able to reproduce the properties of irregularly shaped samples and the departure from spherical symmetry is controlled by the maximum filling factor $0 \leq f_\mathrm{max} < 1$, which we set to $0.8$ (\citealp{Min2016}). The SPF are computed using the \texttt{optool} (\citealp{optool}) using the ``DIANA'' standard opacities (\citealp{Woitke2016}, a mixture of pyroxene and carbon, with a mass ratio of 87 and 13\% and a porosity of 25\%, optical constants from \citealp{Dorschner1995} and \citealp{Zubko1996}, respectively), at a wavelength of $0.8$\,$\mu$m. 

As we can see in Fig.\ref{profspf}, the SB profile is significantly affected by using a more realistic SPF prescription. 
The most striking effect is to be found in the innermost part of the halo, right beyond the PB ring, where the luminosity drop is significantly reduced and even almost completely vanishes at low values of $\theta$. 
In addition, the profile also gets slightly shallower further out in the halo. For the lowest $\theta=15$deg run, the slope is $\sim -3.12$ for the nominal $f_d=8\times10^{-4}$ case (instead of -3.42 for isotrropic scattering) and $\sim-2.95$ for the bright $f_d=4\times10^{-3}$ disc.
This is due to two concurring effects. Firstly, since the average size of bound grains in the halo decreases with stellar distance, asymptotically tending towards $s\sim s_{blow}$ (see Equ.\ref{equ:domsize}), and since, at all scattering angles $\theta$, the considered SPF increases with decreasing grain size in the $s>s_{blow}$ domain (Fig.\ref{DHS}), the relative contribution of outer halo regions to the SB is increased. Secondly, for $\theta\geq4$deg., the SPF actually peaks in the $s<s_{blow}$ domain. This enhances the contribution of unbound grains to the flux (Fig.\ref{unbcurve}) which, as already noted, tends to further flatten the SB profile. .

\subsection{Edge-on configuration}

\begin{figure}
\includegraphics[scale=0.5]{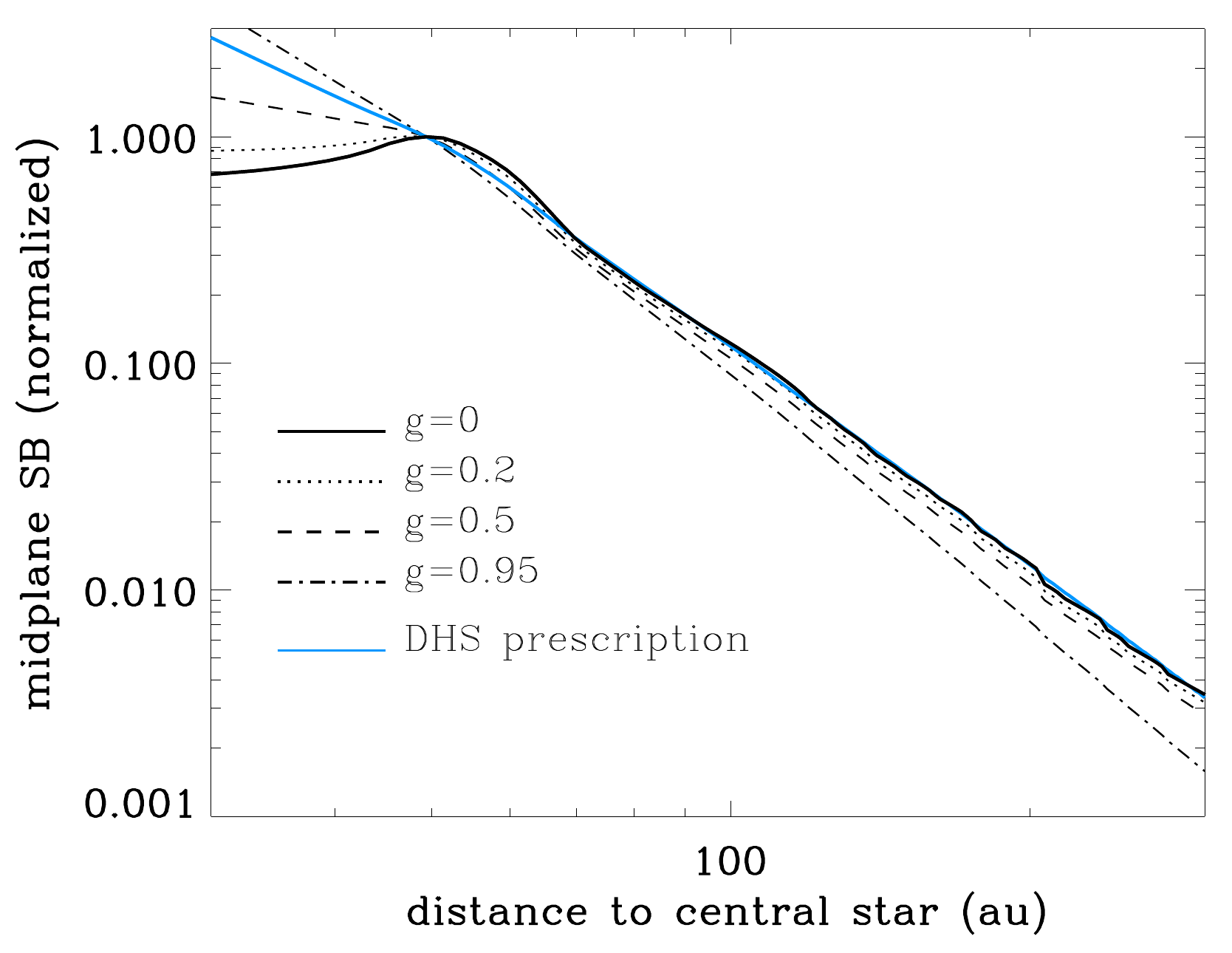}
\caption[]{Edge-on seen disc. Projected radial profile of the midplane surface brightness for 4 different values of the $g$ parameter of the Henyey–Greenstein phase function, as well as for the DHS prescription for the SPF (see main body of the text) }
\label{gcomp}
\end{figure}

\begin{figure}
\includegraphics[scale=0.5]{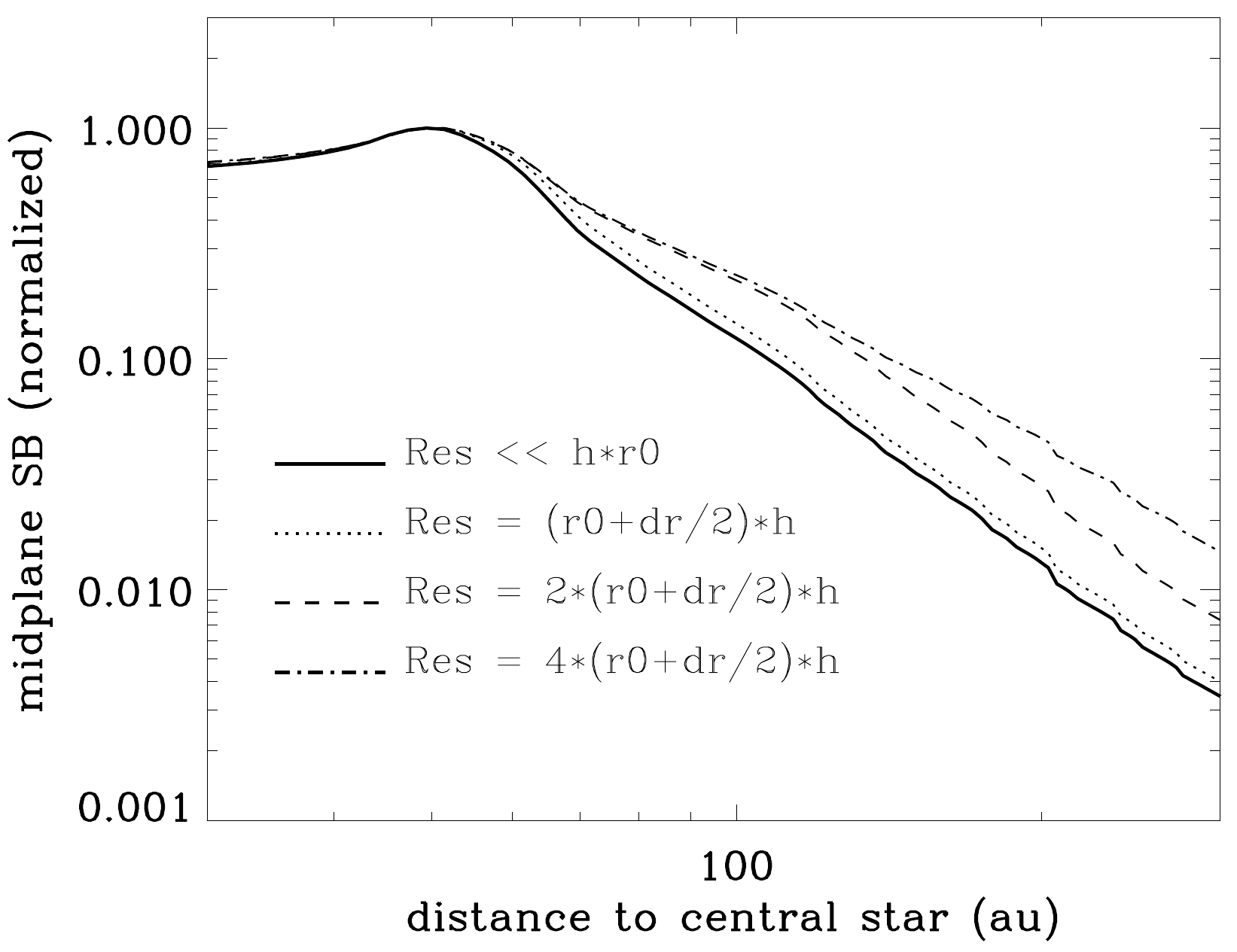}
\caption[]{Edge-on seen disc. Projected radial profile of the midplane surface brightness for 4 different values of the instrument's resolution. $h$ is the disc's aspect ratio, $r_0$ the centre of the parent body belt and $dr/2$ its width (isotropic scattering is assumed)}
\label{rescomp}
\end{figure}

The edge-on case introduces additional specific issues that did not affect the general configuration presented in the previous section. The first one regards the scattering phase function. The SB is indeed now measured in the disc's midplane along the projected distance $\rho$, and is thus now the sum of contributions coming from different physical stellocentric distances $r$, for which the scattering angle is not the same. As a result, the resulting luminosity should in principle depend on the SPF even for SPF prescriptions that do $not$ depend on grain size, and hence stellar distance. We test the importance of this effect by considering the classical Henyey-Greenstein prescription, for which the anisotropy of the scattering behavior of the dust grains is characterized by the dimensionless asymmetry parameter $-1 < g < 1$:
\begin{equation}
    f(\theta) = \frac{1-g^2}{4\pi\left(1+g^2-2g \cos(\theta)\right)^{3/2}}.
\end{equation}
The $g=0$ case represents isotropic scattering, in which photons are scattered in all directions with equal probabilities. For positive (negative) values of $g$, incident photons are scattered in the forward (backward) direction, and as $g$ increases (decreases) the asymmetry is even more pronounced. In practice, the value of the asymmetry parameter $g$ is a way to control the size of the particles, since grains much smaller than the wavelength of observations are expected to scatter isotropically (hence $g=0$), while grains larger than the wavelength should display a strong forward-scattering peak ($g >0$).

Fig.\ref{gcomp} presents the projected midplane profiles \emph{SB($\rho$)} for the isotropic ($g=0$) case as well as for 3 different values of $g$, assuming that the disc is not flared and has a constant aspect ratio $h=H/r$. For the isotropic scattering case, we get SB($\rho$) $\propto r^{-3.3}$ in the outer regions, which is close to the standard -3.5 value that directly follows from the aforementioned relation $\Gamma = \gamma_{out}+1+\delta$ for a constant $h$ (i.e., $\delta=1$).
We see that the projected \emph{SB($\rho$)}  profile strongly depends on $g$ in the innermost $\rho<r_0$ regions. This is expected, because the scattering angle for the most luminous grains, i.e., those in the PB ring, can reach very small values at small $\rho$, which leads, for strong forward scattering (i.e., high $g$ values) to a significant increase of \emph{SB($\rho$)} for decreasing $\rho$. In the outer regions beyond the projected outer edge of the PB belt, the dependence on $g$ is weaker but still noticeable. For the most extreme $g=0.95$ case, the radial index reaches $\gamma_{out}\sim-3.7$ in the outer regions, which is $\sim-0.4$ steeper than for the isotropic ($g=0$) case. This weaker dependence is due to the fact that, in these outer regions, the variation of scattering angles with $\rho$ are more limited than in low $\rho$ regions.
Interestingly, taking a more realistic and size-dependent DHS prescription for the SPF tends to move the SB profile back close to the reference $r^{-3.5}$ slope (blue line in Fig.\ref{gcomp}). This is because of the increased role of small bound and unbound grains that acts to flatten the surface brightness profile (see previous section), and thus acts in the opposite direction of the geometrical effect of integrating along the line of sight.

Another potentially important issue for the edge-on configuration is whether or not the disc is resolved in the vertical direction. The midplane profiles presented in Fig.\ref{gcomp} indeed implicitly assume that the disc is resolved in $z$ at all projected distances $\rho$, so that there is a natural geometrical dilution of the flux in the vertical direction. However, vertical resolution is not necessarily achieved with current instrument facilities (see Sec.\ref{discu}). We explore the effect of not resolving the disk vertically in Fig.\ref{rescomp}, by considering 4 different cases, ranging from fully-resolved-everywhere (resolution $res \ll h \times r_0$) to a very poorly resolved case where only the outermost $>4r_0$ regions are vertically resolved. As can clearly be seen, the \emph{SB($\rho$)} midplane profile flattens with decreasing vertical resolution. For the lowest resolution ($res=4 \times h \times r_0$) case, the slope for the midplane profile tends towards $\gamma_{out}=-2.3$. This value is $\sim \gamma_{out(res=0)}+1$, which is expected because the vertical dilution term is now absent and the "midplane" luminosity does here correspond to the $z$-integrated flux.

\subsection{Wavelength dependence and SED}

\begin{figure}
\includegraphics[scale=0.5]{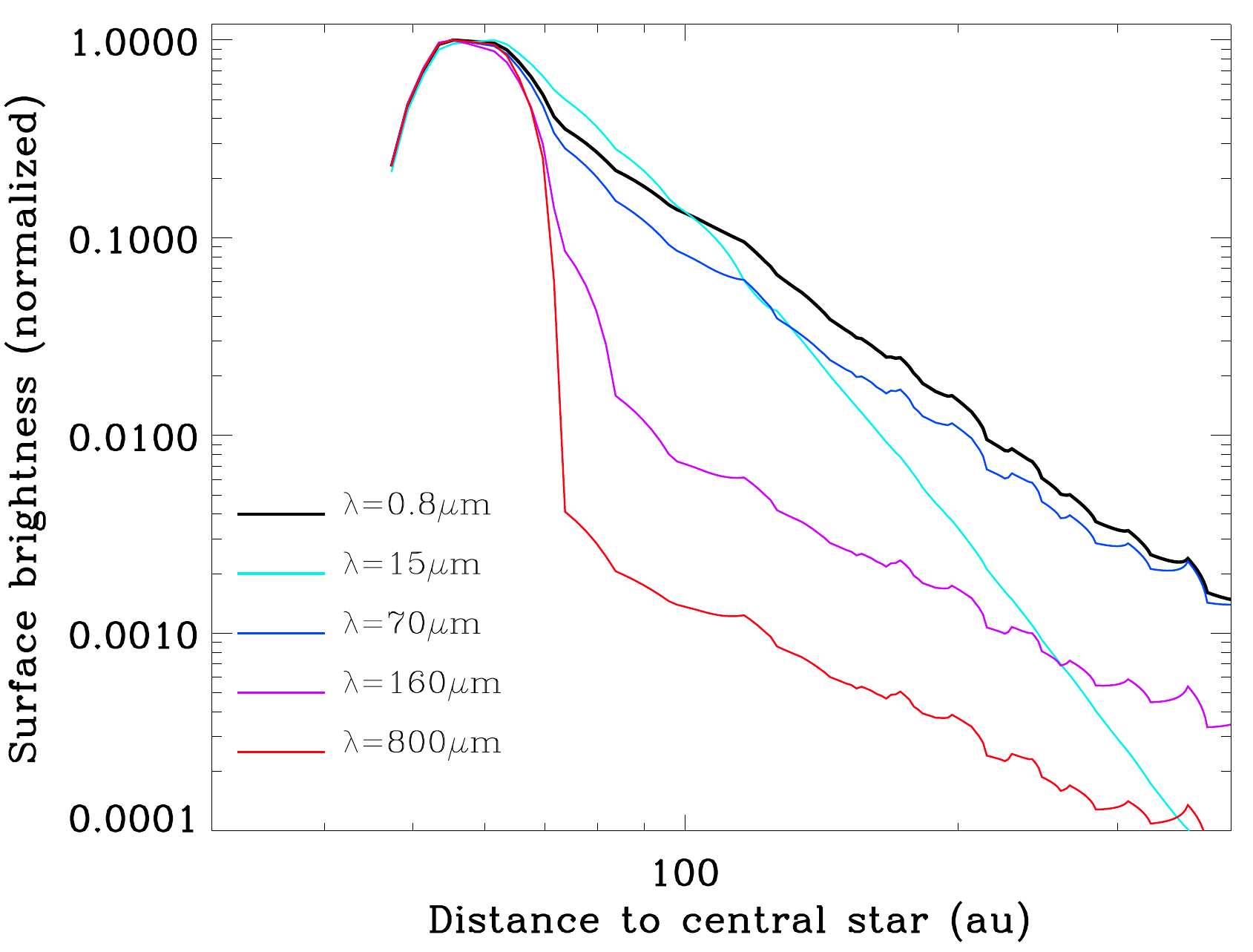}
\caption[]{Radial profile of the normalized surface brightness at four different wavelengths, estimated with the GRaTeR package for the nominal setup}
\label{flcomp}
\end{figure}

\begin{figure}
\includegraphics[scale=0.5]{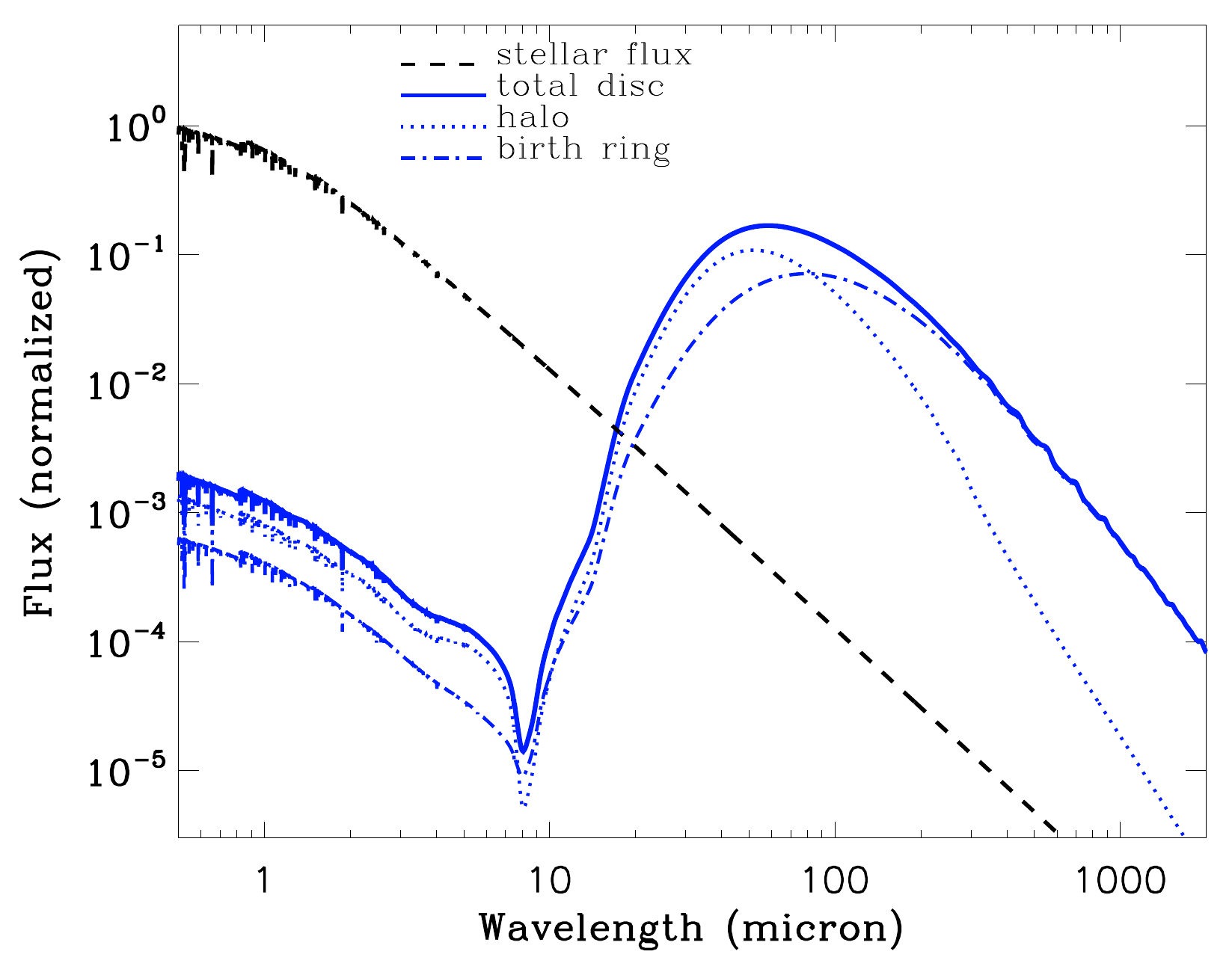}
\caption[]{Normalized system-integrated SED for the nominal setup ($f_d=8\times10^{-4}$), displaying also the respective contributions coming from the parent body belt (between 50 and 66\,au) and the halo (beyond 66\,au). }
\label{narsed}
\end{figure}

\begin{figure}
\includegraphics[scale=0.5]{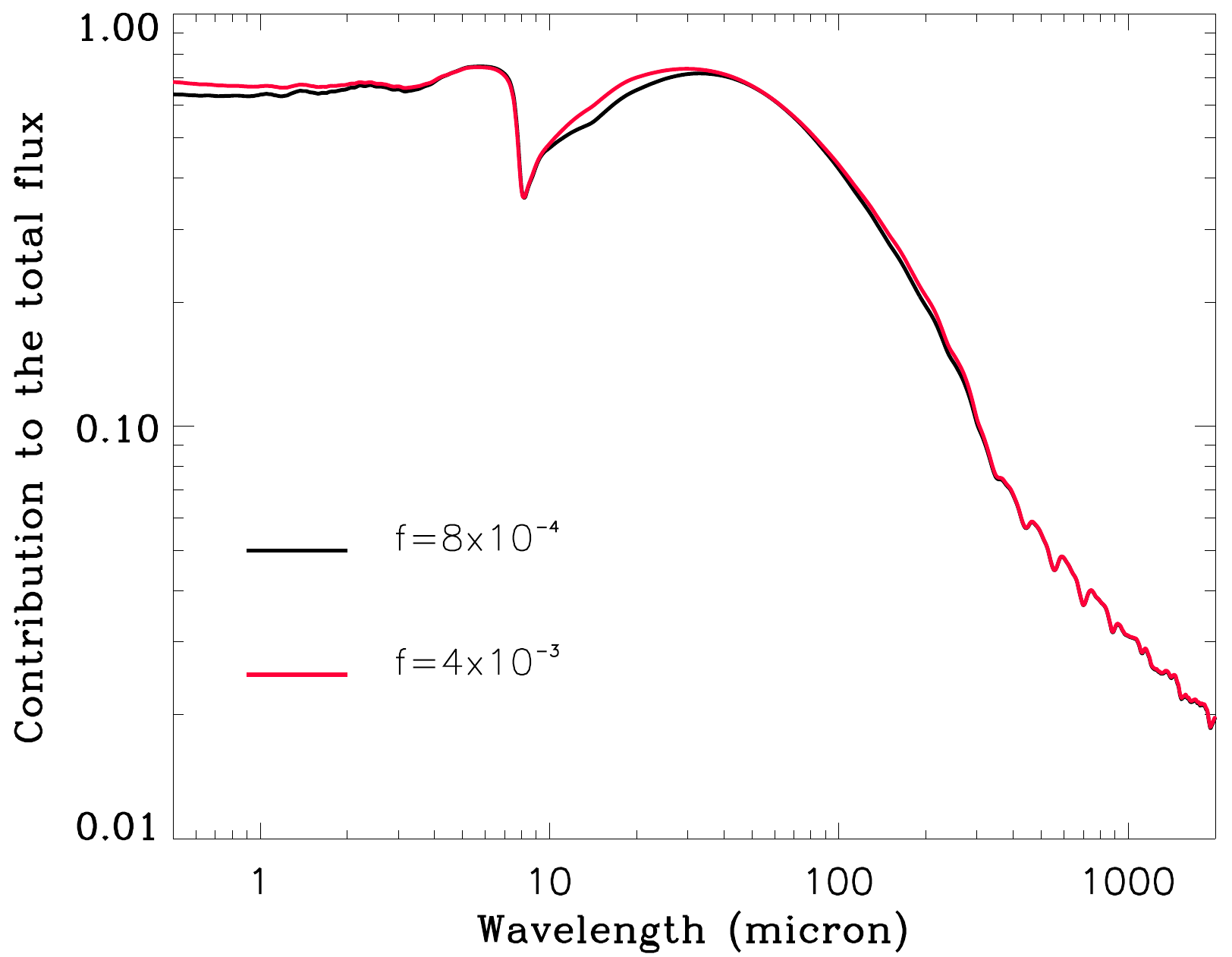}
\caption[]{Relative contribution to the total flux, as a function of wavelength, coming from the whole halo for both the nominal and "very bright" disc cases. The discontinuity at $\lambda\sim8\mu$m corresponds to the transition between the scattered-light dominated domain and the thermal-emission dominated one.}
\label{frach}
\end{figure}

We present here, for the first time, an exploration of the halo phenomenon at over a wide range of wavelengths, notably in thermal emission up to the millimetre domain. 

Fig.\ref{flcomp} presents, for the non-edge-on configuration, how the \emph{SB}(r) profiles vary with $\lambda$. 
At long wavelengths (160 and 800$\mu$m) there is a sharp drop at the outer edge of the parent body belt. This is due to the fact that, at these large $\lambda$, the absorption coefficient $Q_{abs}$ of the small grains that dominate the geometrical cross section is much smaller than 1. As a consequence, the flux within the PB belt is dominated by large $s \gg s_\mathrm{\rm{blow}}$ grains, whose number density abruptly drops beyond the PB belt because their orbits are only very weakly affected by stellar radiation pressure. Beyond $r_0+\Delta r_0/2$, small grains take over, but their higher number density cannot compensate for their very low $Q_{abs} \ll 1$ values. Note that, in this dimmer far-IR and mm halo, the radial profile index of the SB is $\sim -2.2$, which is significantly shallower than in scattered light. The likely interpretation for this is that, at these long wavelengths, the temperature $T$ of grains in the $\sim70-400\,$au regions implies that they emit very close to the Rayleigh-Jeans approximation of the Planck function, for which the flux at a given wavelength $\lambda$ is $\propto T$. Since the blackbody temperature of grains is in turn proportional to $r^{-0.5}$, it follows that the relation between the slope $\Gamma$ of the vertical optical depth $\tau$ and the slope $\gamma_{out}$ of the SB(r) profile is $\Gamma = \gamma_{out} + 0.5$ instead of $\Gamma = \gamma_{out} + 2$ (see Equ.\ref{gamma}). 

At $\lambda=70\mu$m, however, the drop at the outer edge of the PB belt is much more limited, and is only $\sim30\%$ more pronounced than in scattered light. This is because, at this wavelength, the poorly-emitting (i.e., with $Q_{abs}<1$) grains come from a narrow size range $s_{blow}\leq s \leq \lambda/2\pi\sim10\mu$m that only accounts for $\sim60\%$ of the total cross section in the birth ring \footnote{For a standard particle size-distribution in $s^{-3.5}$ and neglecting the contribution of $s<s_{blow}$ unbound grains}. Moreover, even grains in this $s_{blow}\leq s \leq \lambda/2\pi$ domain still have non-negligible $Q_{abs}$ values at $\lambda=70\mu$m \citep[typically $\sim0.2-0.3$, e.g.,][]{morales2013}. As for the slope of the profile, it is $\sim-2.9$, slightly steeper than at longer wavelengths, which results from the fact that the Rayleigh-Jeans approximation becomes less accurate at these shorter $\lambda$. 

The SB profile at $\lambda=15\mu$m is very different, with an absence of luminosity drop at the edge of the PB belt, followed by a very steep decrease with radial distance in the halo. This is because, at this wavelength and at $r\geq60\,$AU from an A6V star, we are in the Wien side of the Planck function, for which the flux increases exponentially with $T$. As a consequence, small grains, even unbound ones, whose temperatures exceeds the almost black-body temperature of larger particles, totally dominate the flux \footnote{their higher T being enough to compensate for their lower emissivity \citep[for a detailed discussion on this issue, see][]{theb2019}} , and there is no drop at the outer edge of the PB ring due to the sudden absence of large grains.
However, because of the exponential dependence of the flux on T in the Wien domain, the decrease of the SB with radial distance becomes steeper and steeper as the grains get colder in the outer halo, going from a radial index of $\sim-4.3$ just outside the PB belt to $\sim-6$ in the $300-400\,$au region.

In Fig.\ref{narsed} we look at the wavelength dependence from the perspective of the system-integrated Spectral Energy Distribution. We see that, despite corresponding to grains that are further away from the star, the halo's SED actually peaks at a shorter wavelength than the PB belt's contribution. In addition, at all wavelengths shorter than $\sim90\mu$m, the relative contribution from the halo to the total flux $F_{halo}/F_{disc}$ exceeds $50\%$ and thus dominates that of the PB ring (Fig.\ref{frach}). These results agree well with those of the radial profiles, showing that there is no sharp luminosity drop at the PB belt/halo interface for wavelengths short of 70$\mu$m.
The only exception is a narrow wavelength range around $\lambda\sim8-15\mu$m. This corresponds to a "sweet spot" where the thermal flux dominates scattered light but is in the Wien regime of the Planck function, for which there is a very steep decrease of the flux with radial distance (see above) and the halo's contribution is thus much lower. Note, however, that the $\lambda\sim10\mu$m domain corresponds to wavelengths at which the disc is very faint (Fig.\ref{narsed}).
At longer wavelengths ($\geq90\mu$m), we logically see a decrease of the halo's contribution to the total flux, which is the direct consequence of the low emissivity of its small grain population at increasing $\lambda$. Nevertheless, even in the mm-wavelengths domain, the halo's contribution never gets totally negligible. As an example, it is still $5\%$ of the total flux at $\lambda=800\mu$m.  

From Fig.\ref{frach} we also see that, when considering the global $F_{halo}/F_{disc}$ ratio, we obtain relatively similar results for both the nominal and bright-disc cases. This is because, in terms of the relative contribution of the halo, the only parameter distinguishing these two cases (once normalized) is the fraction of unbound grains in the system. The higher fraction of submicron grains for $f_d=4\times10^{-3}$ will only affect the flux at short wavelengths ($\lambda\lesssim1\mu$m), where we see that $F_{halo}/F_{disc}$ is indeed $\sim10\%$ higher for this bright disc case, as well as for the aforementioned $\lambda\sim8-15\mu$m domain, where the system's total flux is  dominated by unbound grains \emph{within} the PB belt \citep[see][]{theb2019}.

\section{Discussion} \label{discu}

\subsection{How universal are the SB $\propto r^{-3.5}$ and $\tau\propto r^{-1.5}$ profiles?}

\begin{figure}
\includegraphics[scale=0.5]{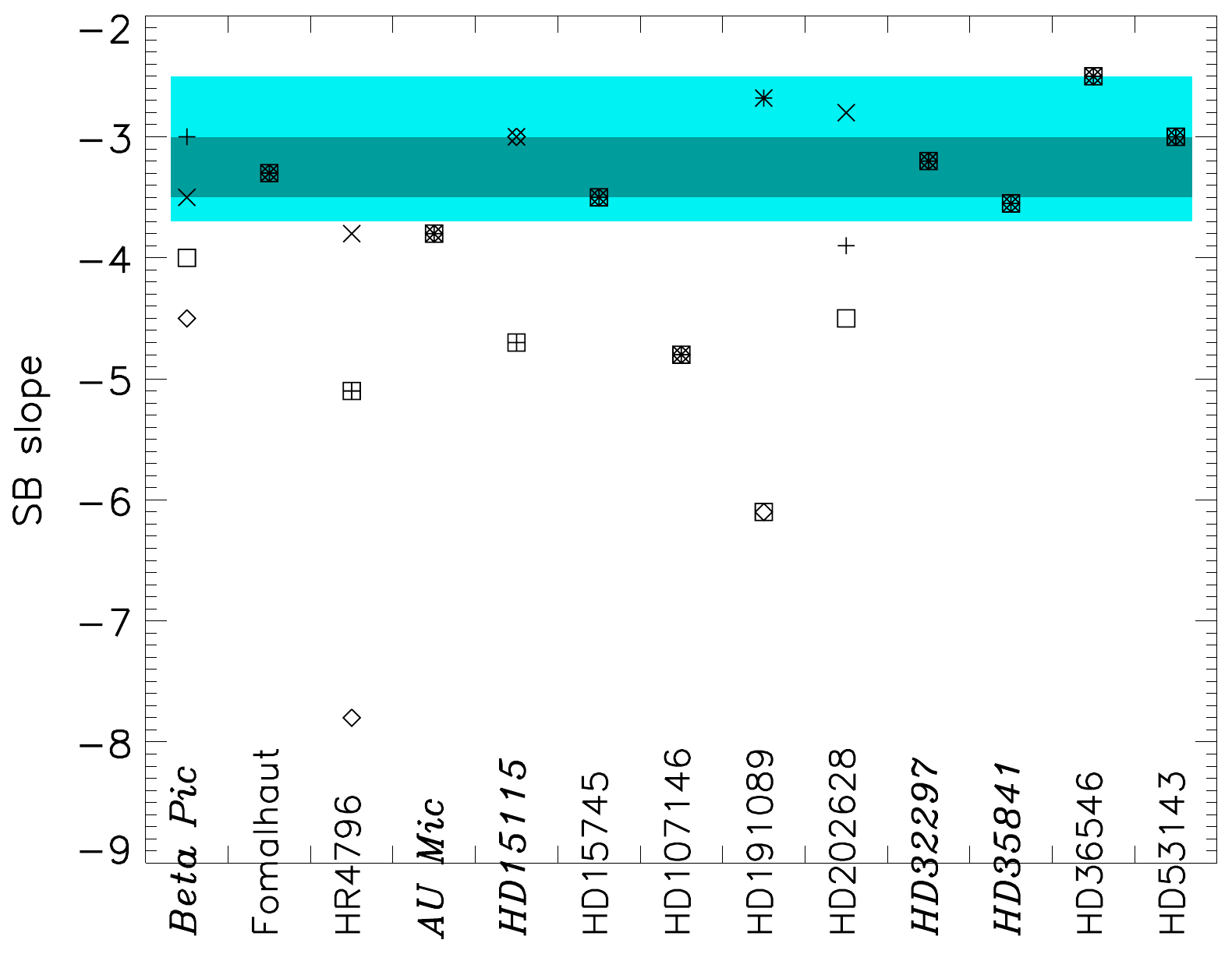}
\caption[]{Surface brightness radial slopes taken from Tab.\ref{tab:slopes}. The dark blue area corresponds to the expected values in a non-perturbed system according to the present numerical investigation. The light blue area is the same, but for edge-on discs, taking into account the potential non-resolution of the disc in the vertical direction. For some systems, there are different slopes estimates depending on the radial position in the halo, as well as depending on which side of the disc has been considered. In this case, up to 4 values are displayed: diamonds and "x" stand for the radial indexes in the inner and outer-halo, respectively, for one disc side, and the squares and "+" are the equivalent indexes for the opposite side. For systems where there is only one global fitted radial index, all 4 symbols do overlap. Edge-on systems are written in italics}
\label{halodat}
\end{figure}

As mentioned in Sec.\ref{intro}, halo radial profiles are sometimes used as a proxy to constrain the level of "unexpected activity" in the outer regions of debris discs: perturbations by (unseen) planets, effect of companion or passing stars, gas drag, etc... This is usually done by measuring the surface brightness profiles or, more often, the extrapolated underlying dust density distribution are compared to the expected "normal" profiles for unperturbed systems derived by \cite{strubbe2006} and \cite{theb2008}. Such direct comparisons do, however, raise several issues. The first one is that the STCH and TBWU reference radial slopes are only valid under several simplifying assumptions, the main ones being the absence of unbound grains and, for the SB profile, isotropic scattering. Conversely, most observation-based fits of the $\sigma$ and $n$ profiles also make some strong simplifications, notably that the grain size distribution is the same everywhere in the system, which is clearly not the case for halos that are on the contrary extremely size-segregated. While this simplification is of limited consequences as long as isotropic scattering is assumed \footnote{note, however, that, in this case, the estimated slopes do not correspond to density profiles but to that of the vertical optical depth $\tau$, see  Sec.\ref{intro}}, it becomes problematic when considering more realistic, and, in particular, size-dependent SPFs. 

We have here, for the first time, numerically explored how these different approximations and simplifications could bias our understanding of debris disc halos. One important result regards the proportion of unbound grains, which always account for more than $10\%$ of the scattered-light luminosity and even dominate the flux in the outer-halo regions for a bright, collisionally active disc (Fig.\ref{unbcurve}). For this $f_d=4\times10^{-3}$ case, the presence of unbound grains is able to significantly flatten both the $\tau$ and SB profiles, since these small particles have a shallower radial distribution in $\tau\propto r^{-1}$. The effect of small grains becomes even more pronounced when considering a realistic size-dependent SPF function, which, for all scattering angles $\theta\geq 4$° always peaks in the $s<s_{blow}$ domain. To a first approximation, the combined effect of high collisional activity and size-dependent SPF results in SB slopes that tend towards $\sim -3$ instead of the standard -3.5 value. Conversely, we expect GRaTeR-type fits of the underlying dust density, which do not take into account these effects, to underestimate the index of the $\sigma$ or $n$ slopes by up to a value of $\sim0.5$.

For the specific case of edge-on discs, this flattening effect, due to the increased influence of unbound grains and of the smallest bound ones, is less visible for the midplane SB profile. This is mainly because it is in large parts compensated by the purely geometrical effect that comes from the fact that the flux at a projected distance $\rho$ is now the sum of contributions integrated along the line of sight, for which the flux dependence on $\rho$ becomes weaker with increasing projected distance. There is, however, a parameter that potentially has a much greater influence on the midplane profile of edge-on discs, which is the non-resolution of the disc in the vertical direction. Our results indeed show that the consequence of not vertically resolving a disc could lead to a flattening of up to +1 in terms of radial index (with an index of -2.5 instead of -3.5) of the SB radial profile of its halo. 
This result might prove important because, even if second-generation instruments, such as SPHERE or GPI, provide a pixel scale of about $12-15$ milli-arcsec \citep[e.g.,][]{maire2016}, and the angular resolution of ALMA observations keeps on improving, down to the au scale for the closest systems, only a handful of systems, and the closest ones, have been resolved in the vertical direction $z$. From Table\,3 and Figure\,11 of \cite{olofsson2022a}, we see that, amongst the list of constrained-halo systems of Tab.\ref{tab:slopes}, only $\beta$-Pic, Au Mic, HR4796, HD115600 and HD61005 have been unambiguously resolved in $z$ at near-IR wavelengths.

With these news results in mind, we can take a renewed look at Tab.\ref{tab:slopes}. Our analysis has shown that the most reliable halo slope estimates are likely to be those made directly on the observed SB profiles, because they do rely on fewer model-dependent assumptions (about the presence of unbound grains or the size-dependence of the SPF). As a consequence, we show in Fig.\ref{halodat} the dispersion of radial slopes for all systems for which it is the SB profile that has been fitted from observations. It can be seen that 6 out of the 13 considered systems do have halos with profiles that are fully within the boundaries of acceptable values found by our numerical investigation: Fomalhaut, AU Mic, HD15745, HD32297, HD35841 and HD53143. On the opposite side of the spectrum, three discs, HR4796, HD107146 and HD36546\footnote{HD36546 falls within the light-blue area but it is not an edge-on system} fall fully outside these boundaries and should thus be systems for which additional mechanisms are sculpting the outer realms of the discs. For the remaining 4 systems, some parts of the halo do have expected radial slopes while some do not, and an assessment of these systems is not possible without undertaking system-specific investigations that go beyond the scope of the present paper. Note, however, that for 2 of these systems, HD191089 and HD202628, the steeper slopes have been measured in a narrow region just outside the PB belt (see the difference between the $r_0$ and $r_{max}$ values in Tab.\ref{tab:slopes}), for which our simulations have shown that a steeper SB profile is in fact expected.

\begin{figure}
\includegraphics[scale=0.5]{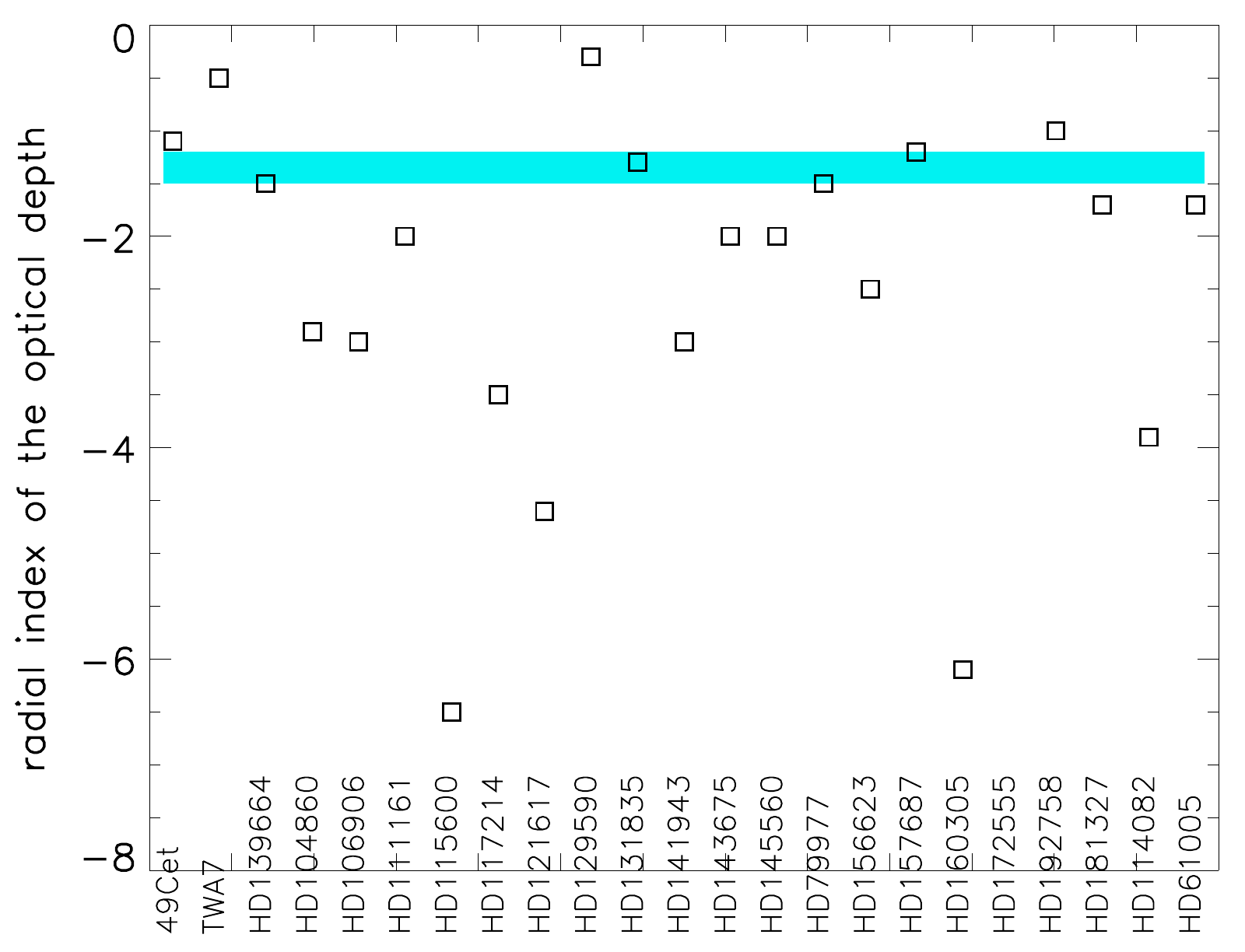}
\caption[]{Values of the vertical optical depth ($\tau$) radial profiles derived from the $n$ and $\sigma$ fits displayed in Tab.\ref{tab:slopes}, when making the simplifying assumption that the radial dependence of $\tau$ is the same as that of $\sigma$, or is equal to that of $n$ plus one (constant opening angle).
The blue domain corresponds to the range of $\tau(r)$ indexes between the -1.48 value obtained for our nominal case and the -1.21 value for our bright disc ($f_d=4\times10^{-3}$) case (see Fig.\ref{optflux}) }
\label{gratergraph}
\end{figure}

As discussed in Sec.\ref{results}, fits of the underlying $n$ and $\sigma$ profiles obtained with GRaTeR or similar codes should be less reliable than fits of the SB, as well as more challenging to interpret in terms of what the fitted quantities mean physically, especially when considering the fact that there is a strong size segregation in the halo. We can, however, consider that the radial dependence of the estimated $\sigma$ and $n$ profiles\footnote{adding one to the $n$ radial slope} should, to a first approximation, give a rough estimate of the radial dependence of the vertical optical depth in these system's halos that can be compared to our results regarding the $\tau$ profiles. Making this approximation of an equivalence between the $\sigma$ (or $n\times r$) and $\tau$ profiles we see in Fig.\ref{gratergraph} that a little less than half of the systems have density radial profiles that are within $\pm50\%$ of the reference $\tau$ profiles obtained in Fig.\ref{optflux}. The spread of radial indexes is larger than for the fitted SB slopes displayed in Fig.\ref{halodat}, with $\sim25\%$ of systems having radial slopes more than 2 indexes below that of our reference simulations. We note, however, that for the 3 systems with the steepest estimated slopes, HD172555, HD160305 and HD115600, the halo profiles are only derived for a relatively limited radial region. As already pointed out, for this narrow region just beyond the PB belt both the SB and $\tau$ profiles are expected to be steeper, so that the slope indexes in this region should not be representative of the halo profile further out. In addition, for some systems where both the SB and $n$ or $\sigma$ radial behaviour have been fitted, we see some incoherent results, with for instance the $\sigma$ profiles being \emph{steeper} than that of the SB for HD15115 and HD32297. This points towards an intrinsic potential pitfall in global ring+halo fits, which we discuss in the next subsection.

\subsection{Suggested procedure for fitting discs with halos}
\label{sec:halofit}

Studies deriving $\sigma$ or $n$ profiles from observations treat the ring+halo system as a whole, usually assuming that the density profiles follows
\begin{equation}
    n(r) = \left[\left(\frac{r}{r_0}\right)^{-2\alpha_{in}}+\left(\frac{r}{r_0}\right)^{-2\alpha_{out}}\right]^{-1/2}n_0
\label{equ:grater}    
\end{equation}
and exploring different values of $r_0$, $\alpha_{in}$ and $\alpha_{out}$, as well as of the disc inclination $i$ and, often, the $g$ parameter of the HG scattering phase function. The best fit is then found by comparison to the observed luminosity profile through a classical $\chi^2$ minimizing procedure. A potential problem is that, because the flux in the bright PB belt is usually much higher than in the halo, the $\chi^2$ fit is dominated by the narrow bright ring, so that acceptable global fits can actually be a poor match of the luminosity profile in the halo region. This could explain the fact that, in some cases, fitted $n$ profiles do seem to disagree with the observed SB profile in the halo (see previous section). More generally, this could lead to large errors when trying to assess the specific structure of the halo.

To alleviate these potential problems we suggest that, in future observational studies, the fitting of the system's radial profile should be done in two steps, in which the main ring and the halo are fitted separately. For non edge-on cases, the PB belt alone could first be fitted by finding the best possible set of $i$, $r_0$ and $\Delta r_0$ values, where $\Delta r_0$ is the FWHM of the ring. Once these parameters are constrained, the halo profile beyond $r_0+0.5\Delta r_0$ can then be investigated by fitting the $\alpha_{out}$ index. 
Such a procedure should ideally take into account the varying size distribution within the halo, the potential effect of unbound grains and the size-dependence of the SPF, all parameters that could potentially change the correspondence between an observed SB profile and the radial profile of the underlying optical depth distribution. Nevertheless, thoroughly exploring these parameters would probably render the fitting procedure much too cumbersome. A possible compromise could be the "semi-dynamical" model experimented by \cite{pawellek2019}, which takes as a starting point a PB belt as constrained from observations in the sub-mm, from which smaller grains are produced with their abundances scaled up by the corrective factor found by STCH and TBWU. Synthetic images are then produced, potentially taking into account realistic SPFs, which are compared to observations in scattered-light. This is, however, not a fit \emph{per se}, even though it provides important information on as to whether or not the observed ring+halo system behaves according to the predictions for unperturbed systems found by STCH and TBWU.
We here propose to approach the problem the other way around, taking as a starting point the constrained SB($r$) profile in the halo, and "reverse engineer" the $\tau(r)$ profile from it. Of course, since the SPF in the halo depends on the grain size distribution, which in turn depends on $r$ and is not known beforehand, there is in principle one unknown too many. However, an approximation of the SPF($r$) dependence could be obtained assuming that, at any given position $r$ in the halo, the geometrical cross section should be dominated by grains of a size $s$, such as their apoastron is located at $r$ when produced at $r_0$ in the main PB belt (Equ.\ref{equ:domsize}). The SPF at position $r$ can then be estimated for this specific grain size $s$ and inputted into the $\alpha_{out}$ fitting procedure.

Note that, in principle, sophisticated numerical codes such as ACE, LIDT-DD \citep{kral2013}, or even the collisional model presented in Section 2 of the present paper, can provide self-consistent estimates of the particle size distribution as a function of radial location in the system that are more accurate than what would be obtained by the procedure suggested in the present paragraph. However, using such sophisticated models to do parameter-best-fits of observed discs would require performing a huge set of CPU-consuming simulations exploring an extended parameter space, which would imply a considerable amount of effort. Why this is undertaken in some rare cases, such as \cite{muller2010} for Vega or \cite{lohne2012} for HD207129, it can only be done within the frame of long and specially dedicated numerical studies. 
What we are aiming for here is different, that is, a relatively easy and "quick-and-not-so-dirty" way of constraining main disc parameters without the inherent flaws of only using Equ.\ref{equ:grater} coupled to a radiation transfer code. The procedure we are presenting is typically to be used at the end of observational studies presenting new resolved data.

\subsection{Halos in thermal emission}

Our numerical exploration has shown that halos, despite being made of small micron-sized grains, should remain relatively bright deep into the mid-IR and even far-IR domain. As an example, for a typical debris ring located at 60au, the halo to PB belt brightness ratio at $70\mu$m is still close to what it is in scattered light (Fig.\ref{flcomp}). It is only at wavelengths longer than $\sim100\mu$m that the halo luminosity drops. Note also that, at these long wavelengths, the radial profile (beyond the sharp drop at the PB belt's outer edge) of the SB in the halo is significantly shallower than in scattered light ($\propto r^{-2.2}$ instead of $\propto r^{-3.5}$). 
Another important result of the present study regards halo total fluxes. We have shown that, except for a narrow domain around $\lambda\sim10\mu$m, halos contribute between 50 and 70\% of the system's total flux at all wavelengths short of $\sim90\mu$m. And even in the millimetre-domain the halo still has an integrated luminosity that amounts to a few percents of that of the main parent belt. Perhaps more tellingly, we find that the total, wavelength-integrated thermal emission of the halo is approximately half that of the whole system. 

Our results show that, in thermal emission, the optimal wavelength window for observing halos beyond belts in the $50-70$\,au region is $\lambda\sim20-70\mu$m. This domain covers two bands (24 and 70$\mu$m) of the MIPS instrument on the Spitzer telescope as well the 70$\mu$m band of the PACS instrument on the Herschel telescope. While the resolution and sensibility of Spitzer probably was too limited to explore halos, the situation was more favourable for Herschel-PACS at $70\mu$m. However, most PACS-resolved discs had estimated sizes smaller than the FWHM of the instrument (5.6" at 70 $\mu$m) and were inferred from image-deconvolution \citep[e.g.,][]{Booth2013}, preventing to derive reliable radial profiles. Such radial profiles were obtained for only a very few systems, such as Vega \citep{sibthorpe2010}, $\epsilon$ Eridani \citep{greaves2014}, Fomalhaut \citep{acke2012} or HD207129 \citep{lohne2012}, but without constraining the slopes of the SB profile in the outer regions. A re-investigation of these systems' Herschel data, and in particular an estimate of their SB$(r)$ radial slopes would definitely improve our understanding of the halo phenomenon by comparison with our present results.
In most cases, however, the absence of radial profiles means that, if halos were present, they could not be identified as a fading extension of a bright belt but would anyway have contributed to the estimated disc size. It is thus likely that, at 70$\mu$m, a significant fraction of PACS-derived disc radii correspond to a blend between the main collisional belt and the halo, and cannot be reliably used to trace the location of these systems' dust mass reservoirs. This blending effect would be less important for the 100$\mu$m PACS band, but the resolution is here poorer (6.8") and our results show that, even at this wavelength, the halo still contributes to $\sim40$\% of the system's flux (see Fig.\ref{frach}).
New generations of far-IR instruments would here be crucially needed to untangle the belt and halo contributions by providing reliable radial surface brightness profiles in the $20-70\mu$m domain.
Note that the JWST's longest wavelength of observation, 28.3$\mu$m, does overlap with the low end of our optimal window for halo observations. Given that instrument's unparalleled sensitivity, we do thus expect it to provide us with the first resolved images of debris disc halos in the mid-IR.

Our results also challenge the way disc SEDs are sometimes used to constrain the global particle size distribution (PSD) in resolved systems. The procedure to constrain the PSD is indeed usually to consider the geometrical profile constrained from image-fitting and then find the PSD's power law index $q$ that best fits the SED, under the assumption that the $dN\propto s^{q}ds$ PSD holds everywhere in the system \citep[e.g.,][]{pawellek2019}. However, this procedure is not adapted to systems for which a large fraction of the SED is due to the halo, which has a PSD that is very different from that of the PB belt. In fact, the very notion of a single $dN\propto s^{q}ds$ law  for the halo makes little physical sense, since this region is strongly size-segregated. The PSD for the halo should thus in principle be derived as a function of radial location, but this would imply having reliable estimates of the SED at different locations in the outer regions, which is generally impossible to obtain. A possible intermediate solution would be, as for estimating the SPF, to assume a simplified size-distribution in the halo, where a given radial location $r$ is only populated by monosized grains produced in the PB belt and having their apoastron at $r$ (Equ.\ref{equ:domsize}). With this assumption, and for systems for which the halo's geometry and SB profile have been constrained from image-fitting by the aforementioned procedure, its contribution to the SED can be unequivocally estimated. It can then be subtracted from the total SED to allow estimating the $q$ PSD index in the main belt through the usual procedure. Of course, our simplifying assumption for estimating the halo's SED would imply that the halo's SB is close to $\propto r^{-3.5}$ and should in principle not be used for systems where the SB's profile strongly departs from this radial dependence. However, we believe that this procedure is in any case preferable to a global fit that assumes a uniform PSD everywhere in the system. 
As for the procedure outlined at the end of Sec.\ref{sec:halofit}, sophisticated codes such as ACE or LIDT-DD could be in principle used to obtain more accurate estimates of the SED. However, here again, what we are aiming for is a quick and easy enough procedure that can be used in observation-based studies without the flaws of having to assume a constant PSD everywhere in the system.

The detailed study of individual systems goes beyond the scope of the present paper, but we conclude this section by briefly discussing if the two halo detections obtained with ALMA for HD\,32297 and HD\,61005 \citep{macgregor2018} can be explained by the "natural" behaviour of halos at long wavelengths without invoking additional mechanisms. We first note that, contrary to our numerical results, the radial profiles in the millimetre of these two halos appear relatively steep, with -6.2 and -5.5 for the extrapolated surface density's index $\Gamma$ instead of -1.5. In addition, the radially-integrated deprojected luminosities $F_{halo}$ of these halos amount to between 20 and 30\% that of parent body belt $F_{belt}$, which is one order of magnitude more than for our synthetic halos at $\lambda=1.3\mu$m (see Fig.\ref{frach}). This points towards an "abnormal" halo and thus the need for additional mechanisms at play in the outer regions of these systems \footnote{ Interestingly, \cite{krivov2018} have identified HD61005 as being potentially self-stirred, but it remains to be seen how self-stirring can alter the brightness profiles of outer regions}
Note, however, that the fitting procedure adopted by \cite{macgregor2018} imposes a density continuity at the belt/halo interface and it is not clear to what extent this assumption, coupled to the almost edge-on orientation of the system, affects the obtained results in terms of density slopes and $F_{halo}/F_{belt}$ ratios.

\subsection{Unresolved systems}

Our results have also consequences for systems that have not been resolved at any wavelength and whose size and radial location $r_d$ are only constrained by their SED. One of the most sophisticated method to retrieve $r_d$ from the SED is the one proposed by \cite{pawellek2014} and \cite{pawellek2015}. The first step of this procedure is to fit the SED with a modified black body (MBB) model \citep{backman1993} to derive the typical dust temperature $T_{dust}$, which is in turn used to derive a blackbody radius $r_{BB}$. The "true" disc radius $r_d$, which implicitly corresponds to that of the PB belt, is then obtained by multiplying $r_{BB}$ by a factor $\Gamma_{(L*)}$ that depends on stellar type. The $\Gamma_{(L*)}$ ratio is obtained separately by empirically comparing $r_{BB}$ to real disc radii for a sample of \emph{resolved} systems in the far-IR with Herschel. To their credit, \cite{pawellek2014} were aware of the risk of incorrectly estimating $r_d$ if considering wavelengths at which small, radiation-pressure grains can contribute, and thus chose disc sizes retrieved from Herschel PACS images at $\lambda=100\mu$m instead of $70\mu$m. However, as already mentioned, our results show that, even at this relatively long wavelength, the halo of small grains still makes out $\sim40\%$ of the system's total flux \footnote{ because the Q$_{abs}$ of the grains dominating the cross section is still of the order of 0.2-0.3 at this wavelength, see Section 3.3} and could thus lead to overestimating $r_d$. A more reliable estimate would be obtained by considering ALMA images in the mm domain, but such data was not readily available at the time the $\Gamma_{(L*)}$-based procedures were developed. We thus strongly recommend updating the $\Gamma_{(L*)}$ empirical law taking as a reference $r_d$ values determined from ALMA images, in the spirit of the study by \cite{matra2018} or \cite{pawellek2021}

Note that our results do challenge the notion that a single MBB law can accurately fit a system that is actually made of two distinct components, PB belt and halo, which have very different spatial structures and particle size distributions, and whose SEDs peak at different wavelengths. As a consequence, the "disc temperature" $T_d$ derived by the MBB procedure probably has a limited physical meaning and cannot be a reliable estimate of the temperature in the PB belt. Since the halo's SED peaks at a shorter wavelength than that of the PB belt, $T_d$ probably overestimates the actual temperature of the collisionally active region of the disc. This does not, however, invalidate the $\Gamma$-based procedure, as it is in principle independent of the fact that $T_d$ has a physical meaning or not. What matters here is the reliability of the empirical $\Gamma_{(L*)}$ ratio estimates, for which $T_d$ (or rather $r_{BB}$) can be considered as an abstract proxy.

\section{Summary and conclusion}\label{ccl}

We have carried out the most thorough investigation of the halo phenomenon to date. We focus in particular on two issues: 1) how robust the theoretical $\tau \propto r^{-1.5}$ and $SB \propto r^{-3.5}$ radial profiles are when taking into account the role of unbound grains, realistic SPF prescriptions and instrument resolution. 2) How halos behave in thermal emission, out to the millimetre domain, both on resolved images and on system-integrated SEDs.

For a typical halo produced beyond a collisional belt located at $\sim60\,$au, our main results can be summarized as follows:
\begin{itemize}
    \item The contribution of small unbound grains amounts to at least $\sim10\%$ of halo luminosities in scattered light, and can even dominate in the outer halo regions for bright discs. For these brightest discs, halo radial profiles can become significantly shallower.
    \item Size-dependent scattering phase functions (SPF) do also result in flatter radial profiles, which directly follows from the fact that halos are strongly size-segregated regions.
    \item For edge-on viewed systems, not resolving the disc in the vertical direction can lead to a flattening of the SB$(\rho)$ radial profile index by up to one 
    \item Comparing these new results to a complete sample of observationally-constrained halo SB profiles, we find that roughly half of them have radial profiles fully compatible with our predictions, while $\sim25\%$ have profiles that cannot be explained by our models (being usually too steep). For these systems, additional mechanisms should be at play to shape the outer regions. For a large fraction of the remaining $\sim25\%$, halo profiles have been derived in a too narrow region to allow reaching definitive conclusions.
    \item We obtain comparable results for systems for which it is the underlying dust density distribution whose radial profile has been observationally constrained. However, these density distribution fits should be less reliable than SB ones. They should in particular be biased by the fact that they do not discriminate between belt and halo, and that they do not take into account the effect of unbound grains and size-dependent SPFs. 
    \item We suggest that future observational fits of the underlying density distribution in systems with halos should be made in two steps, starting with a geometrical fit of the PB belt whose parameter are then injected in a separate fit of the halo's radial slope that accounts for size-dependent SPF effects
    \item Radially extended halos should also be visible in thermal emission in the $\lambda\sim20-100\mu$m range, where the halo-to-main-belt contrast is comparable to what it is in scattered light 
    \item With the exception of a narrow $8\lesssim\lambda\lesssim15\mu$m domain, halos always account for more than $50\%$ of the disc's total flux up to $\lambda\sim90\mu$m. 
    \item Despite being located further out than the PB belt, the halo's SED peaks at a shorter wavelength than that of the belt and thus makes the global system appear hotter
    \item Beyond $\lambda\sim90\mu$m, halo brightness strongly decreases with wavelength, but halos still contribute to a few percents of the flux in the millimetre domain. This seems, however, not to be enough to explain the bright halos detected with ALMA around HD32297 and HD61005
    \item For unresolved discs, the presence of a halo can also bias the procedure inferring their radial location from an analysis of their SED. 
\end{itemize}
The study of individual systems goes beyond the scope of the present paper and is deferred to future studies, in which additional parameters, such as stellar type, PB belt radial location and dynamical context (such as known stellar or planetary companions) will be explored.

\begin{acknowledgements}
J.\,O. acknowledges support by ANID, -- Millennium Science Initiative Program -- NCN19\_171.
\end{acknowledgements}

\bibliographystyle{aa}

\clearpage

\end{document}